\documentclass[12pt,a4paper]{iopart}

\usepackage{braket}
\usepackage{amssymb}
\usepackage{stackrel}

\usepackage{graphicx}
\usepackage{subfigure}

\usepackage{hyperref}

\newcommand{\ba}{\begin{eqnarray}}
\newcommand{\ea}{\end{eqnarray}}
\newcommand{\del}{\ensuremath{\partial}}

\begin{document}

\title[Visualising quantum effective action calculations in zero dimensions]{Visualising quantum effective action calculations in\\
zero dimensions}

\author{Peter Millington, Paul M.~Saffin}

\address{School of Physics and Astronomy, University of Nottingham, University Park,\\
Nottingham NG7 2RD, United Kingdom}

\ead{p.millington@nottingham.ac.uk, paul.saffin@nottingham.ac.uk}

\vspace{10pt}
\begin{indented}
\item[]18 September, 2019
\end{indented}

\begin{abstract}
We present an explicit treatment of the two-particle-irreducible (2PI) effective action for a zero-dimensional quantum field theory. The advantage of this simple playground is that we are required to deal only with functions rather than functionals, making complete analytic approximations accessible and full numerical evaluation of the exact result possible. Moreover, it permits us to plot intuitive graphical representations of the behaviour of the effective action, as well as the objects out of which it is built. We illustrate the subtleties of the behaviour of the sources and their convex-conjugate variables, and their relation to the various saddle points of the path integral. With this understood, we describe the convexity of the 2PI effective action and provide a comprehensive explanation of how the Maxwell construction arises in the case of multiple, classically stable saddle points, finding results that are consistent with previous studies of the one-particle-irreducible (1PI) effective action.
\end{abstract}

\submitto{\jpa}

\maketitle

%%%%%%%%%%%%%%%%%%%%%%%%%%%%%%%%%%%%%%%%%%%%%%%%%%%%%
\section{Introduction}
\label{sec:Intro}
%%%%%%%%%%%%%%%%%%%%%%%%%%%%%%%%%%%%%%%%%%%%%%%%%%%%%

The quantum effective action~\cite{Jackiw:1974cv, Cornwall:1974vz} has become a powerful tool in fundamental physics, providing a means to derive the quantum-corrected equations of motion for the $n$-point functions of a quantum field theory. Once embedded within the Schwinger-Keldysh~\cite{Schwinger:1960qe, Keldysh:1964ud} closed-time-path formalism (see also \cite{Jordan:1986ug, Calzetta:1986cq}), it allows the first-principles derivation of systems of quantum Boltzmann equations~\cite{Baym:1961zz} (see also~\cite{Blaizot:2001nr, Berges:2004yj}), allowing us to describe, for instance, the evolution of particle number densities in the early universe, finding applications in leptogenesis (for recent reviews, see \cite{Dev:2017trv, Dev:2017wwc}) and baryogenesis (see, e.g., \cite{Prokopec:2003pj, Morrissey:2012db}). Once extended by the introduction of a cutoff or regulator function, the so-called effective average action can be used to derive the exact renormalisation-group flow equations~\cite{Wetterich:1992yh, Morris:1993qb} that allow us to analyse the phase transitions and fixed points of field theories, having applications both in condensed matter and high-energy physics (for a review, see \cite{Berges:2000ew}).

The aim of this work is to provide a concrete and explicit exposition of the quantum effective action by considering a zero-dimensional quantum field theory, thereby allowing qualitative understanding obtained from truncated results to be compared directly with the exact numerical result for the path integral. In doing so, we will be able to elucidate a number of subtleties of the 2PI effective action in relation to its convexity (see \cite{Alexandre:2012ht, Alexandre:2012hn} and references therein), the correct interpretation of the sources with respect to which the Legendre transforms in its definition are performed and the various $n$-point variables that play a role in its approximate evaluation. In doing so, we confirm the results of \cite{Garbrecht:2015cla}, wherein it was shown that a careful treatment of the sources allows one to move between variants of the 2PI effective action, including the two-point-particle-irreducible (2PPI) effective action~\cite{Verschelde:1992bs}, and to constrain truncations of the effective action so that symmetries are preserved, in similar spirit to the symmetry-improved effective action~\cite{Pilaftsis:2013xna}. In the case of vacuum transitions between radiatively-generated minima (\`{a} la \cite{Coleman:1973jx}, see also \cite{Weinberg:1992ds}), this treatment of the sources allows a self-consistent calculation of the tunnelling rate~\cite{Garbrecht:2015cla, Garbrecht:2015yza}.

The remainder of this article is organised as follows. In section~\ref{sec:2PI}, we review the two-particle-irreducible (2PI) effective action, as applied to a simple zero-dimensional quantum field theory. We discuss the convexity of the 2PI effective action in section~\ref{sec:Convexity}. In section~\ref{sec:SingleSaddle}, we derive the form of the effective action when the path integral is dominated by a single saddle point, before showing how the Cornwall-Jackiw-Tomboulis (CJT) effective action~\cite{Cornwall:1974vz} is recovered in section~\ref{sec:CJT}. We then turn our attention to the case of multiple saddle points in section~\ref{sec:MultiSaddle}, showing explicitly how the Maxwell construction arises. Our concluding remarks are given in section~\ref{sec:Conclusions}.

All figures presented in what follows are calculated for $\hbar=1$. Unless stated otherwise, all analytic results for the effective action are truncated at order $\hbar^2$ and component quantities are truncated at the relevant corresponding order.

%%%%%%%%%%%%%%%%%%%%%%%%%%%%%%%%%%%%%%%%%%%%%%%%%%%%%
\section{The 2PI effective action}\label{sec:2PI}
%%%%%%%%%%%%%%%%%%%%%%%%%%%%%%%%%%%%%%%%%%%%%%%%%%%%%

\begin{figure}
  	\centering
    	\includegraphics[width=60mm]{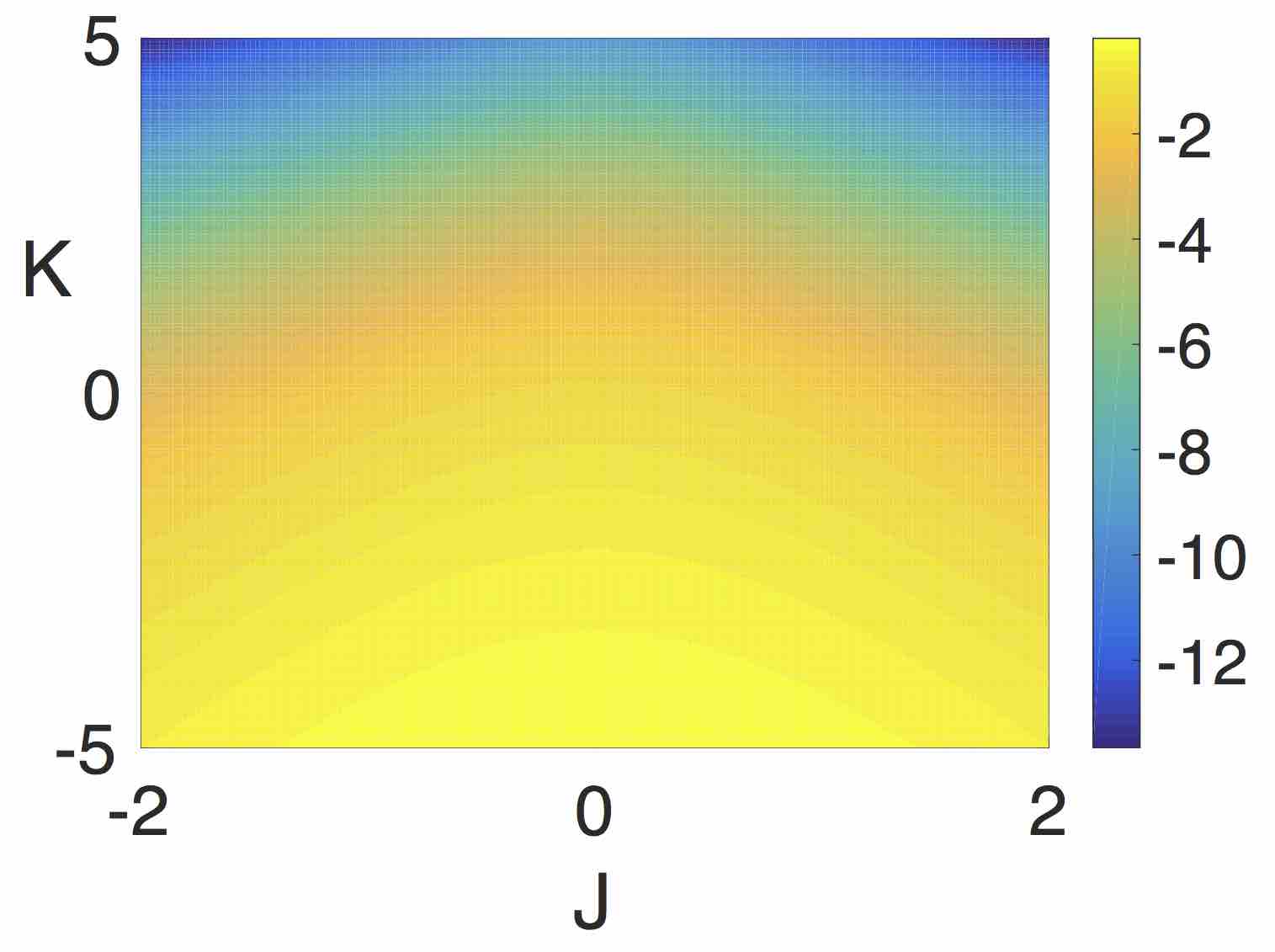}
  	\caption{Plot of $W(J,K)$ for $m^2=-1$ and $\lambda = 6$.}\label{fig:Z}
\end{figure}

We begin by reviewing the definition of the two-particle irreducible (2PI) effective action. We start with the classical action $S(\Phi)$. As a concrete example, we take
\ba\label{eq:action}
S(\Phi)&=&\frac{m^2}{2}\Phi^2+\frac{\lambda}{4!}\Phi^4,
\ea
where $m^2$ and $\lambda$ are real parameters. We can then define the partition function
\ba\label{eq:Z}
Z(J,K)&=&{\cal N}\int^{\infty}_{-\infty}{\rm d}\Phi\;\exp\left[ -\frac{1}{\hbar}\left( S(\Phi)-J\Phi-\frac{1}{2}K\Phi^2 \right) \right],
\ea
where $\mathcal{N}$ is an irrelevant constant normalisation, which we set to unity hereafter, and the sources $J$ and $K$ couple linearly and quadratically to the integration variable $\Phi$, respectively. A plot of the Schwinger function
\ba
W(J,K)=-\hbar \ln [Z(J,K)]
\ea
is shown in figure~\ref{fig:Z}, and we see that it is a concave function of the sources $J$ and $K$. Its first derivative with respect to $-J$ gives the expectation value of the one-point variable in the presence of the sources $J$ and $K$, $\braket{\Phi}_{J,K}$. Its first derivative with respect to $-K/2$ gives the expectation of the two-point variable in the presence of the sources $J$ and $K$, $\braket{\Phi^2}_{J,K}$.

%%%%
%%%%

\begin{figure}
\begin{tabular}{ccc}
  \includegraphics[width=0.3\textwidth]{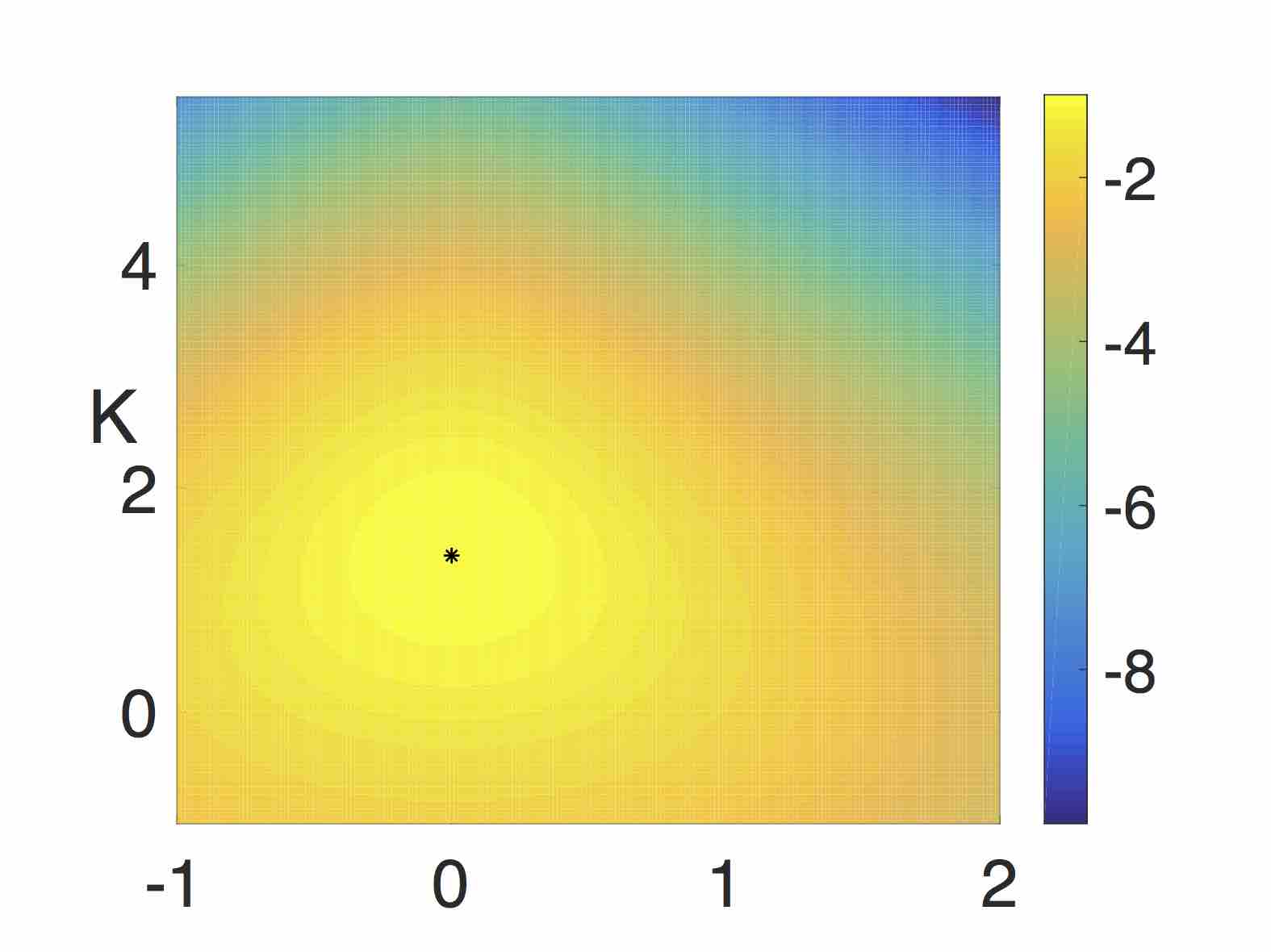} &   \includegraphics[width=0.3\textwidth]{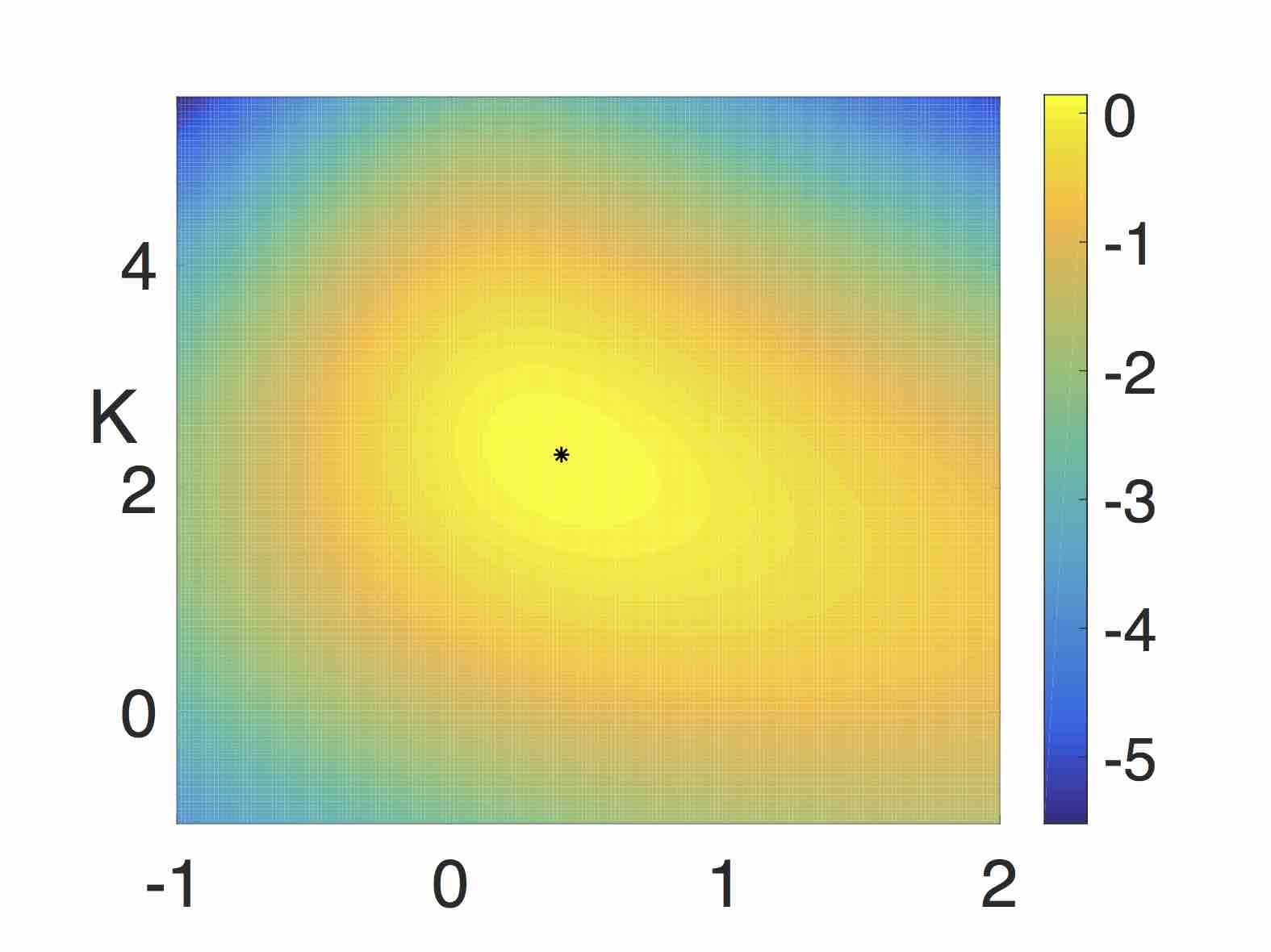} &   \includegraphics[width=0.3\textwidth]{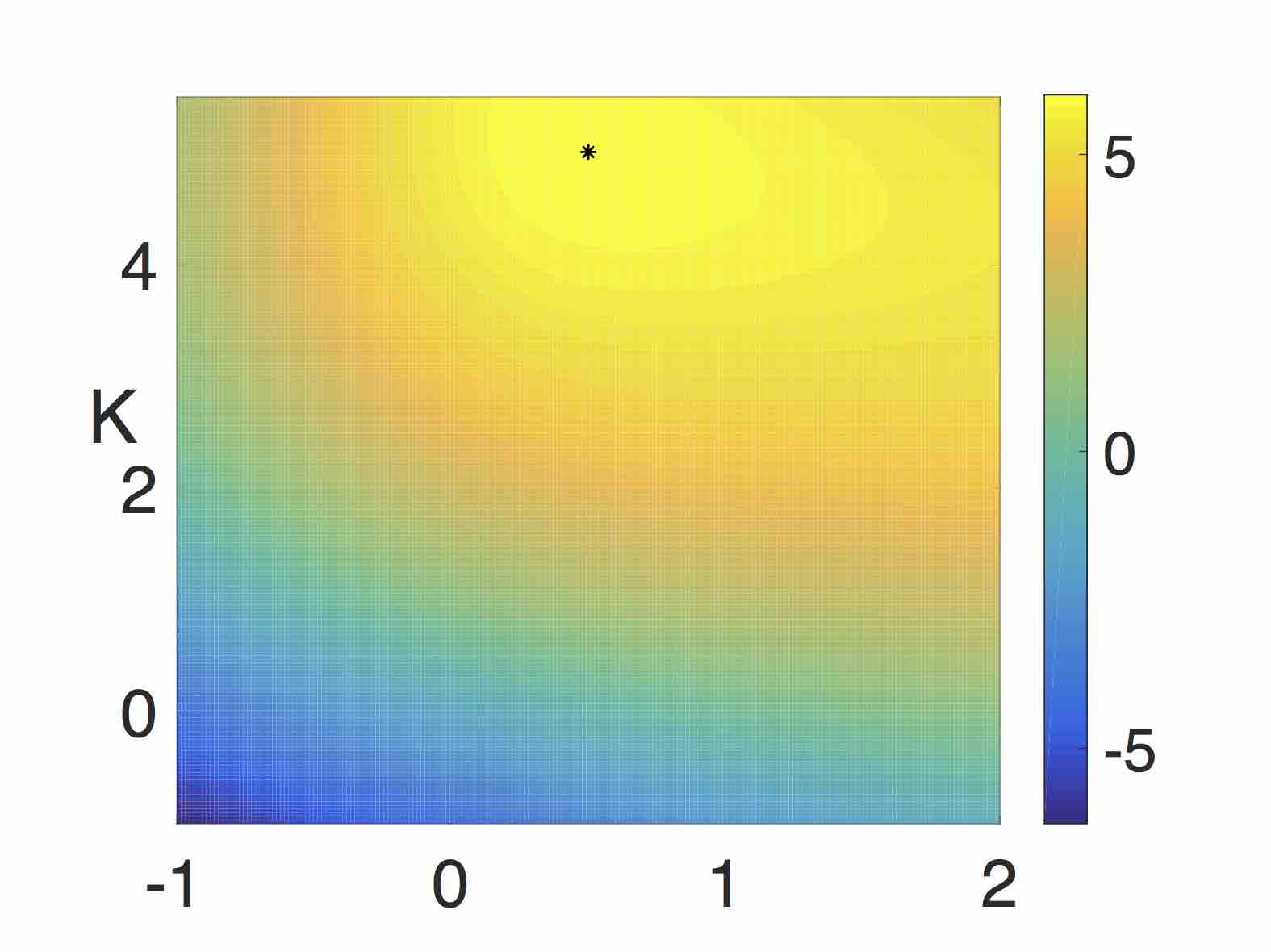}  \\
\small{$\Gamma_{J,K}(0,2)$} & \small{$\Gamma_{J,K}(1,2)$} & \small{$\Gamma_{J,K}(2,2)$} \\[6pt]
   \includegraphics[width=0.3\textwidth]{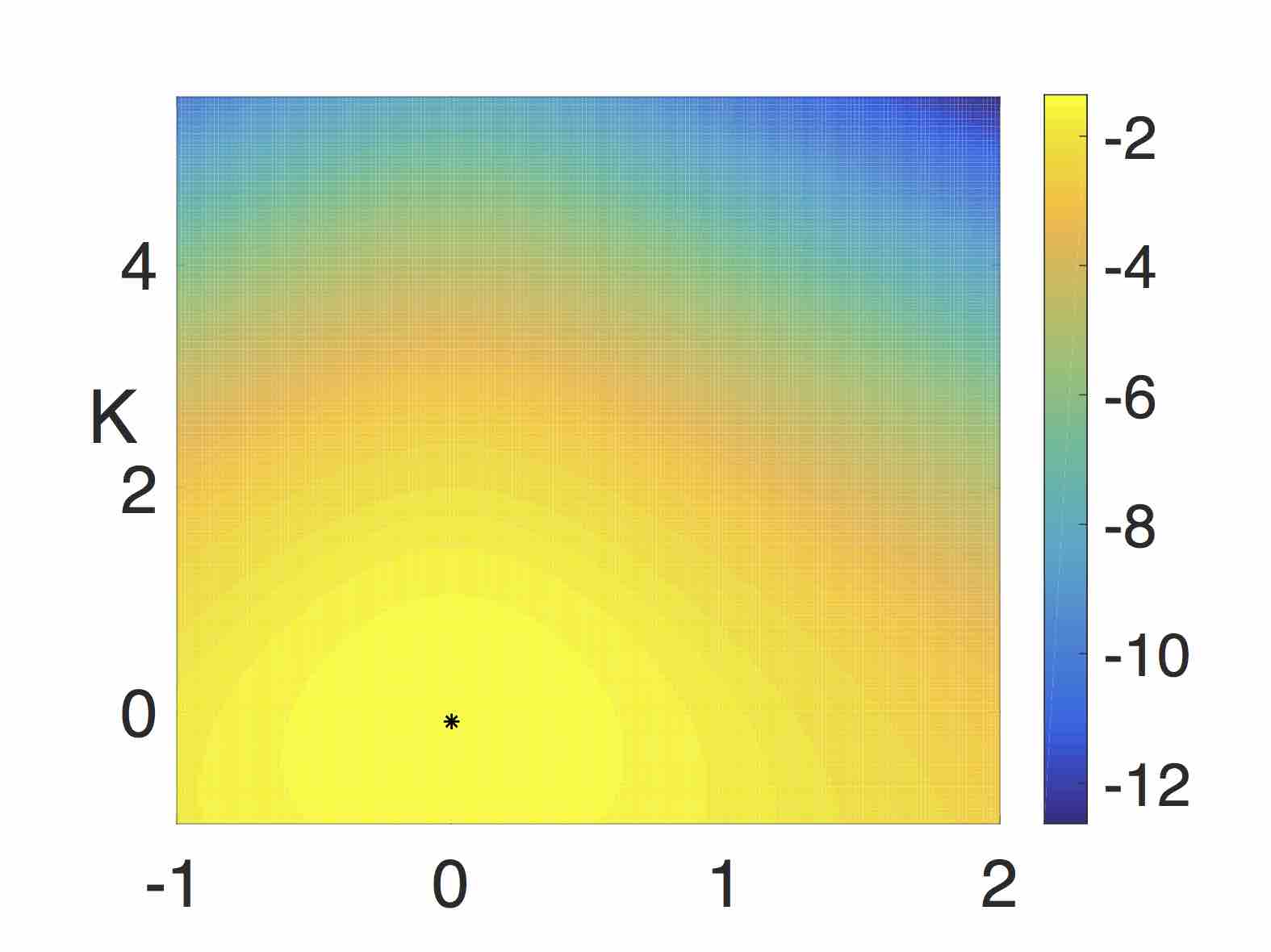} &   \includegraphics[width=0.3\textwidth]{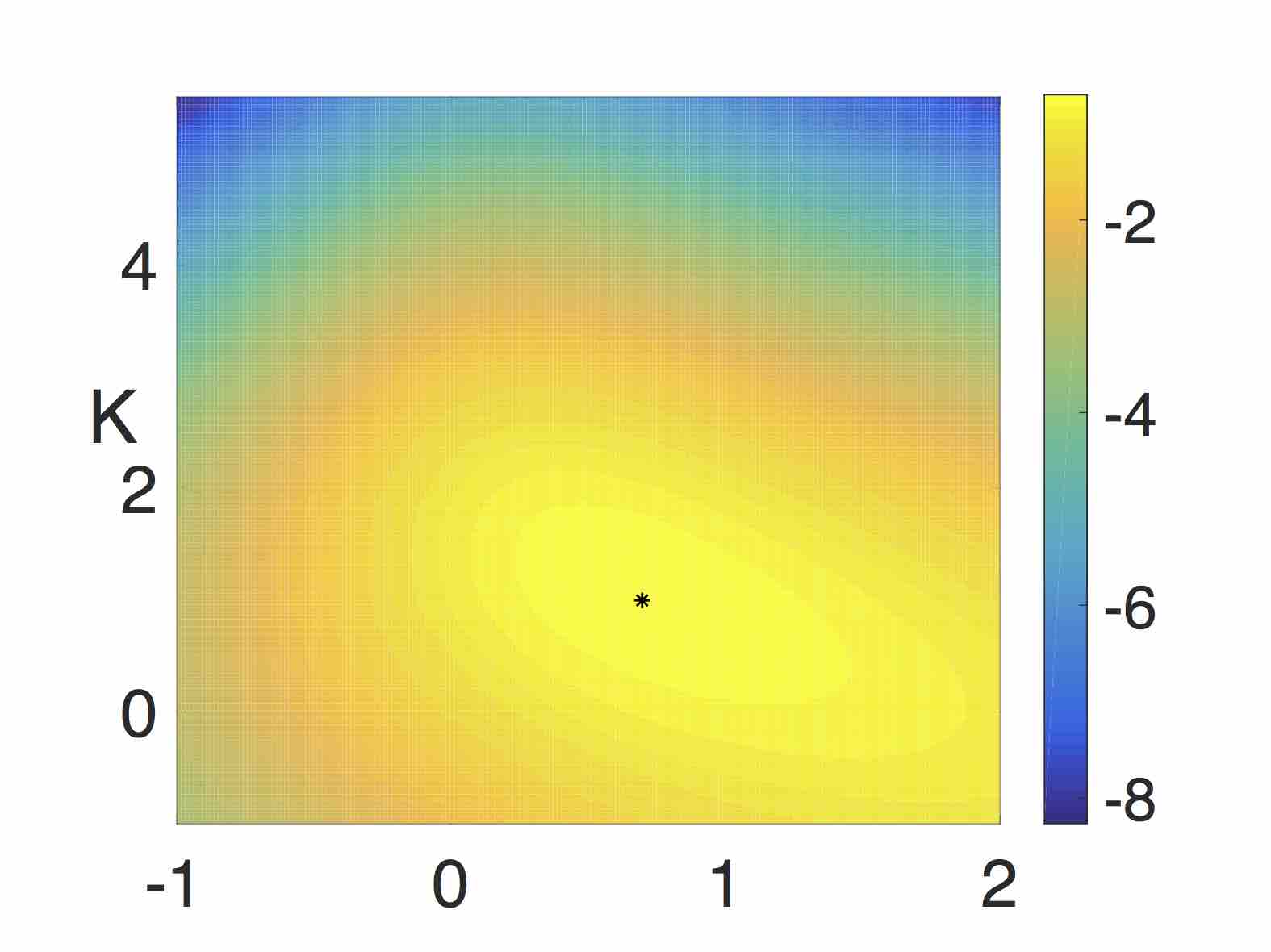} &   \includegraphics[width=0.3\textwidth]{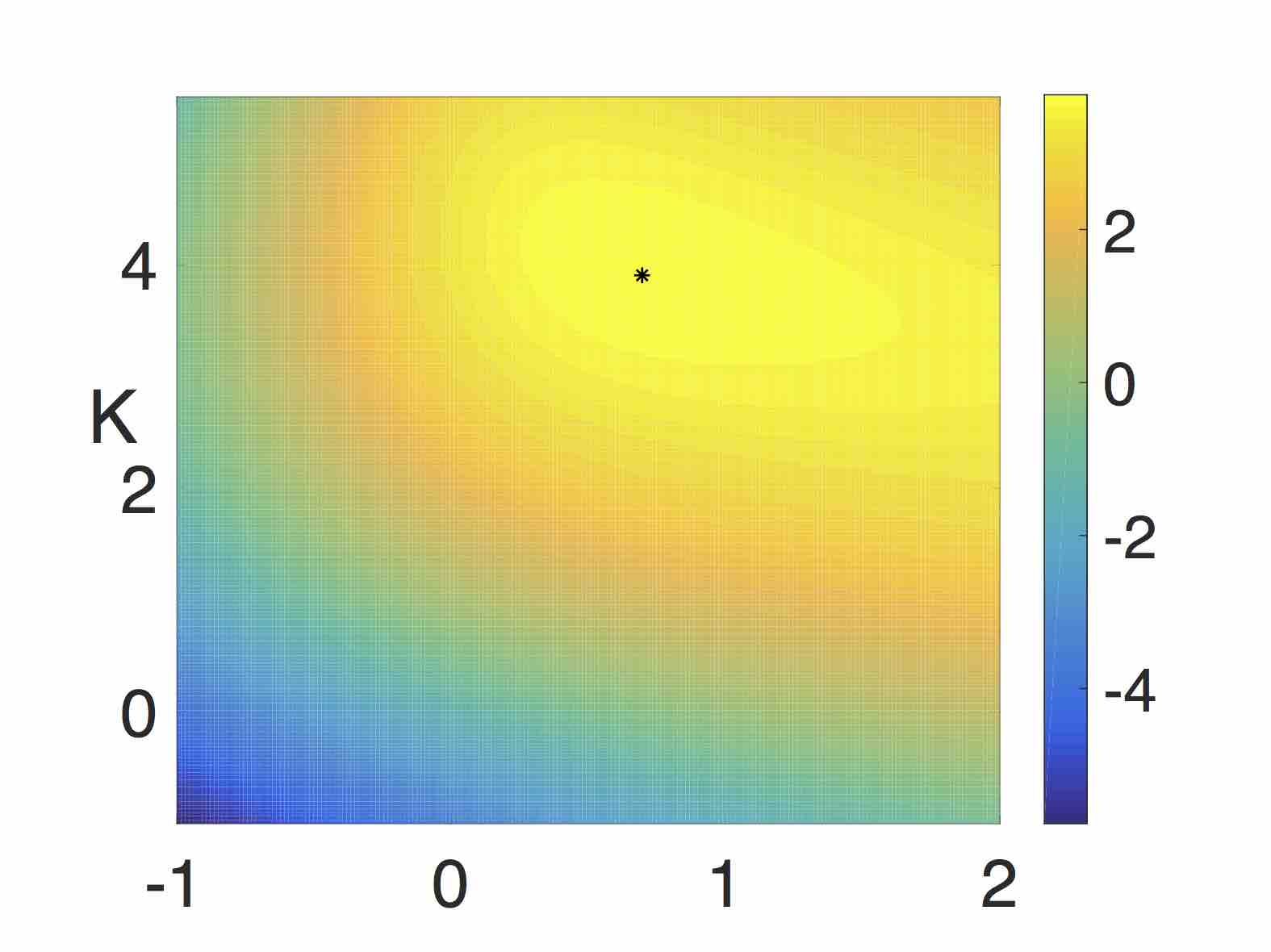}\\
\small{$\Gamma_{J,K}(0,1)$} & \small{$\Gamma_{J,K}(1,1)$} & \small{$\Gamma_{J,K}(2,1)$} \\[6pt]
   \includegraphics[width=0.3\textwidth]{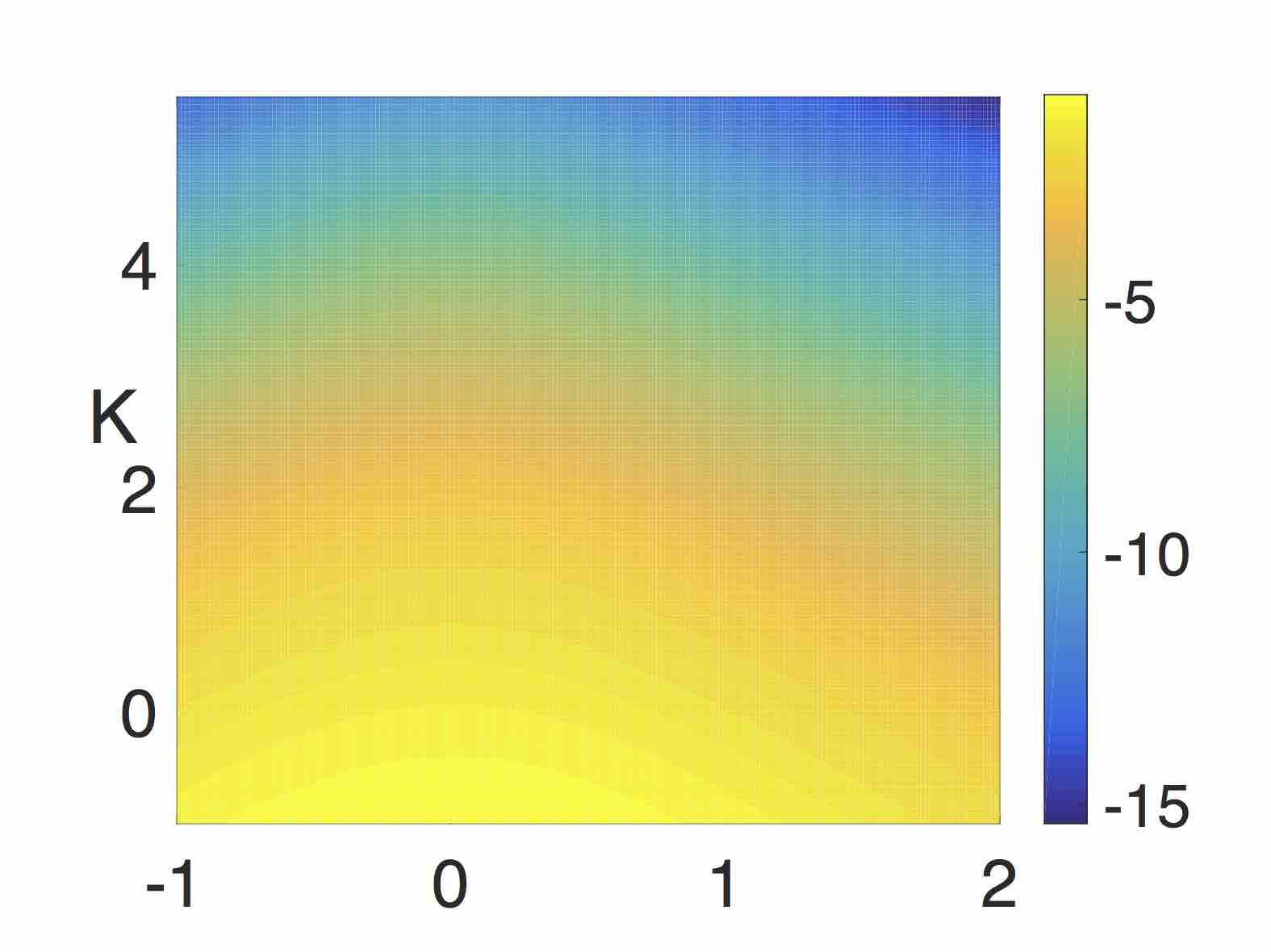} &   \includegraphics[width=0.3\textwidth]{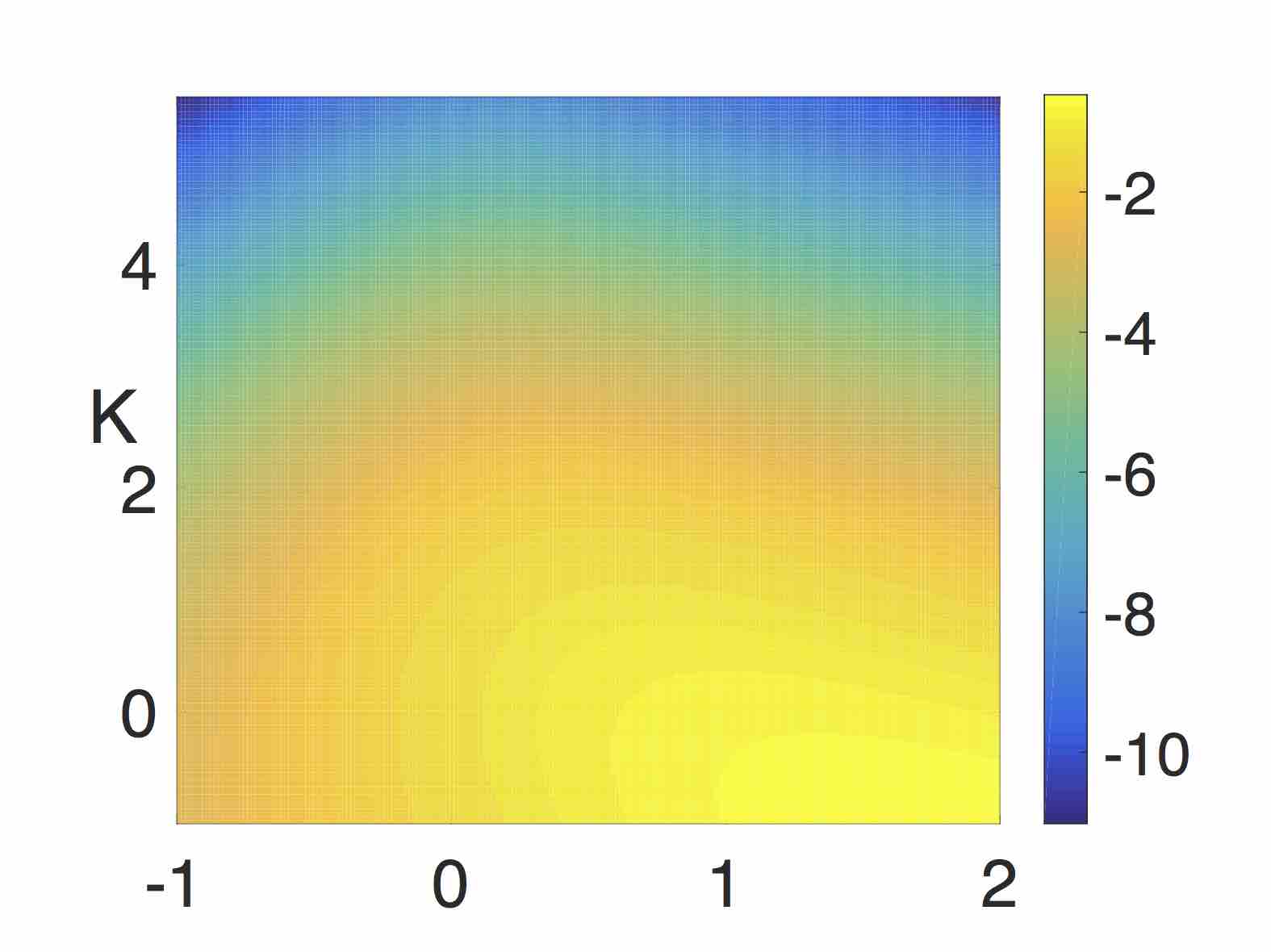} &   \includegraphics[width=0.3\textwidth]{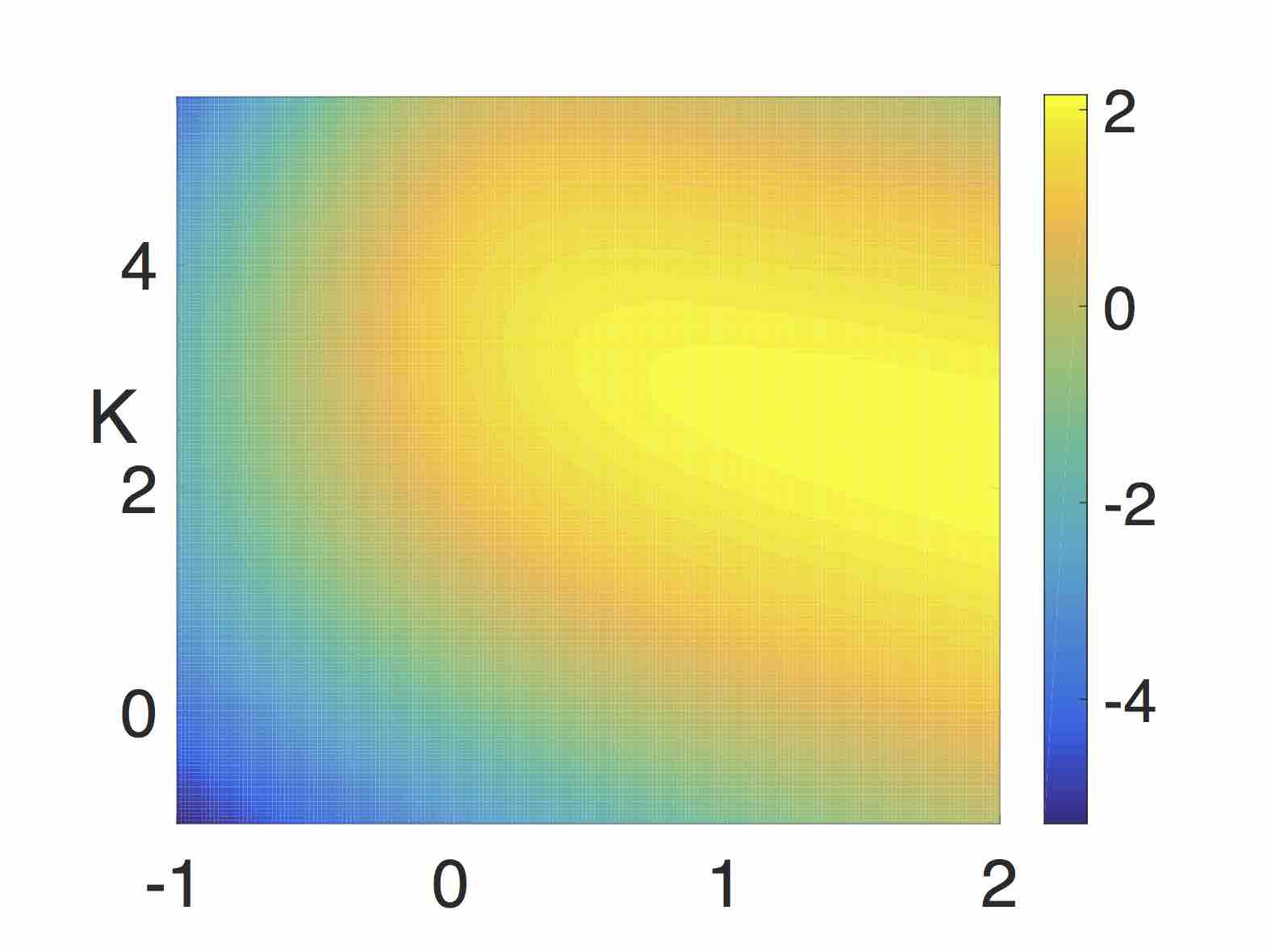}\\
\small{$\Gamma_{J,K}(0,0)$} & \small{$\Gamma_{J,K}(1,0)$} & \small{$\Gamma_{J,K}(2,0)$} \\[6pt]
\end{tabular}
\caption{Plots of $\Gamma_{J,K}(\phi,\Delta)$ for various values of $\phi$ and $\Delta$ for $m^2=-1$ and $\lambda=6$, with the extremum highlighted in each case by a black dot.}\label{fig:Gamma_J_K_phi_delta}
\end{figure}

We now introduce a function that will allow us to define the Legendre transform of the Schwinger function:
\ba
\Gamma_{J,K}(\phi,\Delta)=W(J,K)+J\phi+\frac{1}{2}K[\phi^2+\hbar\Delta],
\ea
examples of which may be seen in figure~\ref{fig:Gamma_J_K_phi_delta} for various values of the variables $\phi$ and $\Delta$. These variables determine the value of the maximum of this function and its position in the $J$-$K$ plane. The Legendre transform
\ba
\Gamma(\phi,\Delta)&=&{\rm max}_{J,K}\Gamma_{J,K}(\phi,\Delta)
\ea
corresponds to the values of these maxima as a function of $\phi$ and $\Delta$, and we denote the locations of the maxima in the $J$-$K$ plane by the extremal sources ${\cal J}$ and ${\cal K}$, defined by
\numparts
\ba
\left.\frac{\del \Gamma_{J,K}(\phi,\Delta)}{\del J}\right|_{J={\cal J}, K={\cal K}}&=&0,\\
\left.\frac{\del \Gamma_{J,K}(\phi,\Delta)}{\del K}\right|_{J={\cal J}, K={\cal K}}&=&0.
\ea
\endnumparts

\begin{figure}
\centering
\subfigure[$\Gamma(\phi,\Delta)$\label{fig:GammaJKa}]{\includegraphics[width=60mm]{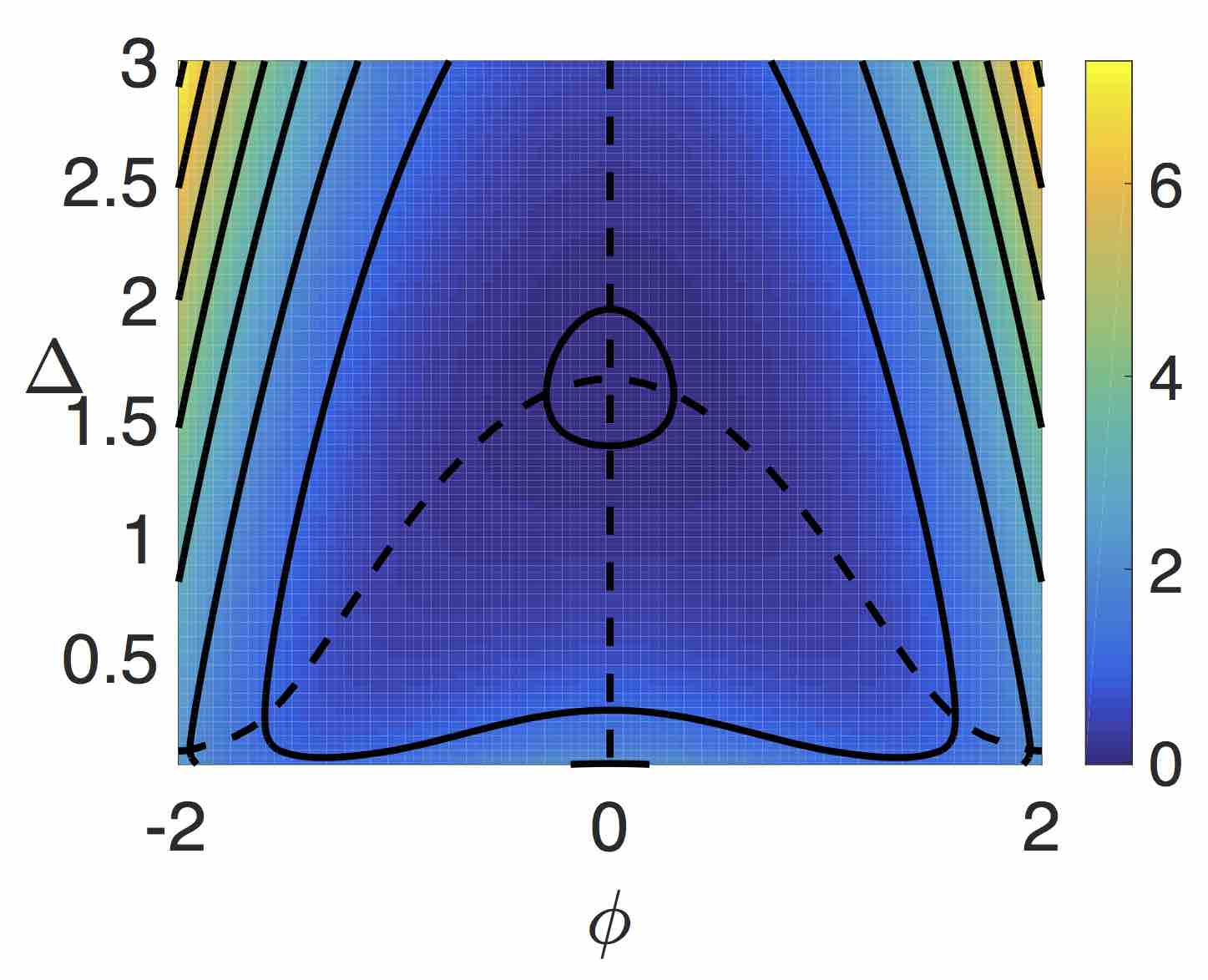}}\\
\subfigure[${\cal J}(\phi,\Delta)$]{\includegraphics[width=60mm]{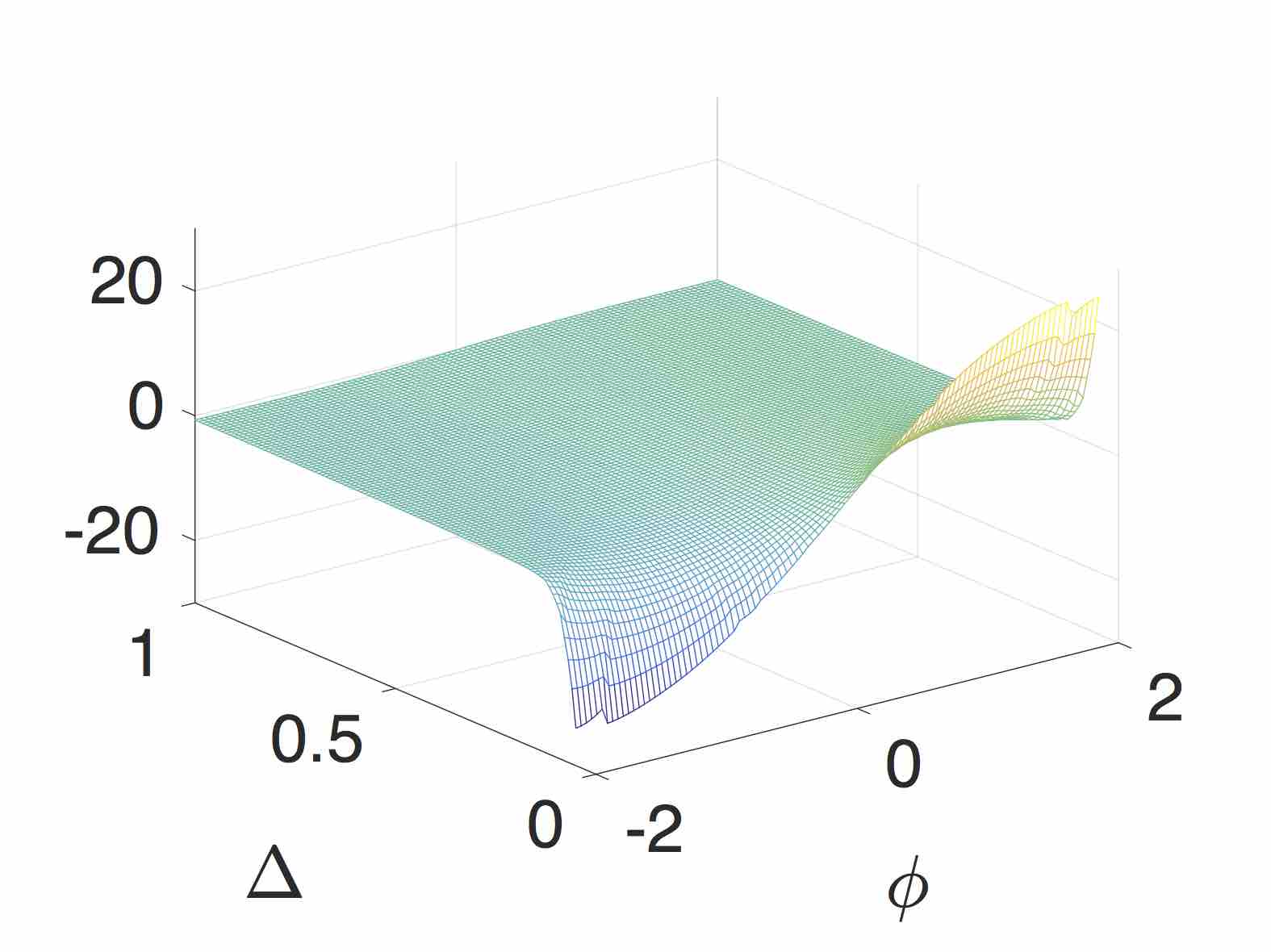}} \qquad
\subfigure[${\cal K}(\phi,\Delta)$]{\includegraphics[width=60mm]{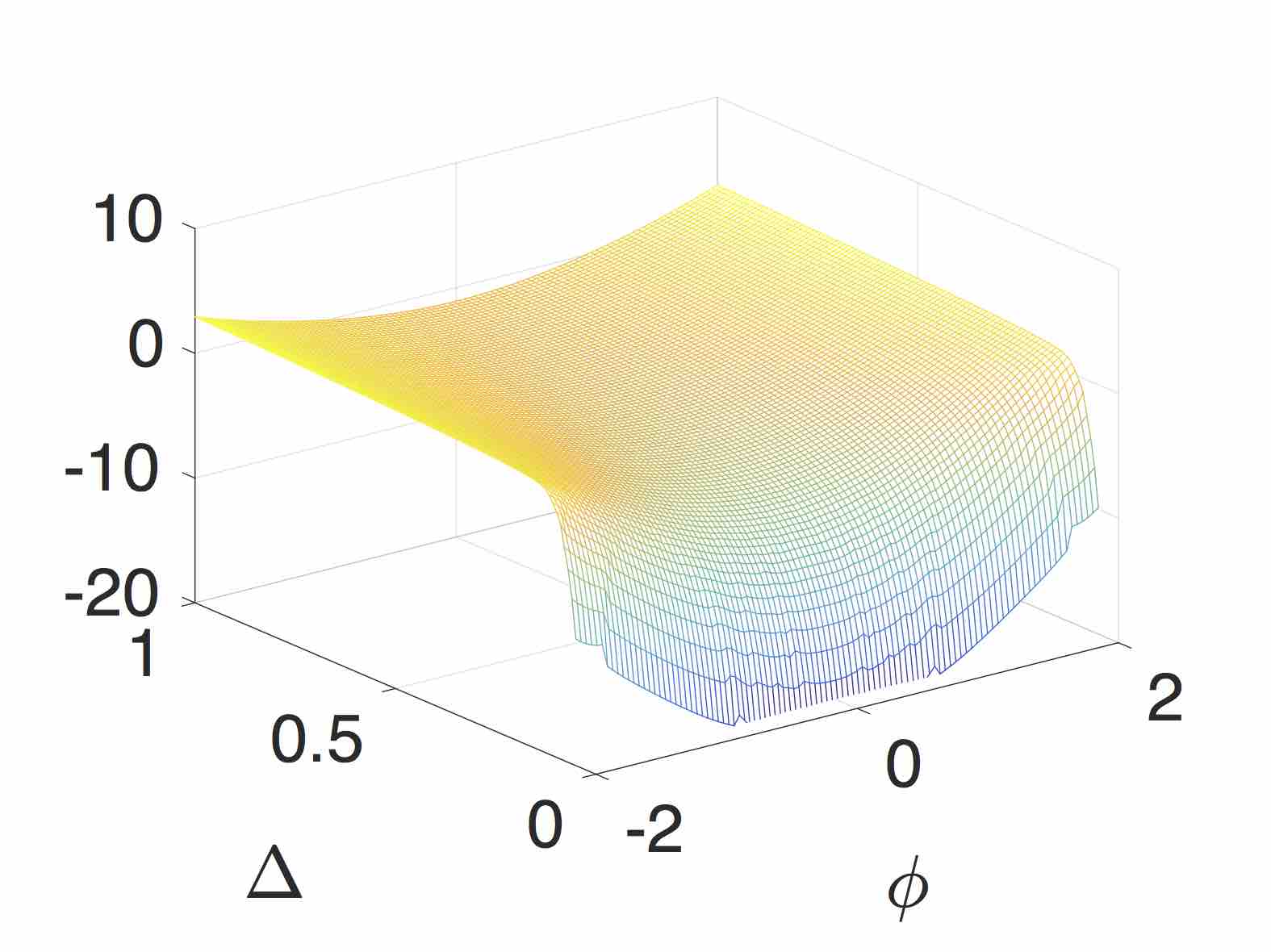}}
\caption{Plots of $\Gamma(\phi,\Delta)$, ${\cal J}(\phi,\Delta)$ and ${\cal K}(\phi,\Delta)$ for $m^2=-2$ and $\lambda=6$ as functions of $\phi$ and $\Delta$. In panel (a), the dashed line from left to right corresponds to the 1PI curve for which ${\cal K}(\phi,\Delta)=0$, and the dashed line from top to bottom corresponds to $\mathcal{J}(\phi,\Delta)=0$. The point where these lines cross corresponds to the extremal point, cf.~section~\ref{sec:CJT}.}\label{fig:GammaJK}
\end{figure}

After performing the extremisation, we obtain
\ba\label{eq:effAction}
\Gamma(\phi,\Delta)&=&W({\cal J}, {\cal K})+{\cal J}\phi+\frac{1}{2}{\cal K}[\phi^2+\hbar\Delta],
\ea
and $\phi$ and $\Delta$ are the connected one- and two-point variables given by
\numparts
\ba
\label{eq:phi}
\phi&=&\hbar\left.\frac{\del}{\del J}\ln(Z)\right|_{J={\cal J}, K={\cal K}},\\
\label{eq:Delta}
\hbar\Delta&=&2\hbar\left.\frac{\del}{\del K}\ln(Z)\right|_{J={\cal J}, K={\cal K}}-\phi^2.
\ea
\endnumparts
We emphasise that, since the location of the maxima of $\Gamma_{J,K}(\phi,\Delta)$ depend on the values of $\phi$ and $\Delta$, we have that
\ba
{\cal J}&\equiv&{\cal J}(\phi,\Delta)\quad {\rm and}\quad {\cal K}\equiv{\cal K}(\phi,\Delta)
\ea
are functions of $\phi$ and $\Delta$. These are plotted in figure~\ref{fig:GammaJK} for the example in (\ref{eq:action}). In corollary, we have that $\phi\equiv \phi(\mathcal{J},\mathcal{K})$ and $\Delta\equiv \Delta(\mathcal{J},\mathcal{K})$. These variables are related to the tangents to the Schwinger function, which can be reconstructed from their envelope. Instead, the extremal sources $\mathcal{J}$ and $\mathcal{K}$ are related to the tangents to $\Gamma(\phi,\Delta)$; namely, it follows from (\ref{eq:phi}) and (\ref{eq:Delta}) that
\numparts
\ba\label{eq:J}
\frac{\del\Gamma(\phi,\Delta)}{\del\phi}&=&{\cal J}(\phi,\Delta) +{\cal K}(\phi,\Delta)\phi,\\\label{eq:K}
\frac{\del\Gamma(\phi,\Delta)}{\del\Delta}&=&\frac{\hbar}{2}{\cal K}(\phi,\Delta).
\ea
\endnumparts
Since the right-hand sides of these expressions are source terms, we see that the gradients of $\Gamma(\phi,\Delta)$ correspond to the equations of motion for the one- and two-point functions. Moreover, these equations of motion contain terms beyond the classical action at all orders in the parameter $\hbar$, and this justifies the naming of $\Gamma(\phi,\Delta)$ as the \emph{quantum effective action}.

%%%%
%%%%

%%%%
%%%%

%%%%%%%%%%%%%%%%%%%%%%%%%%%%%%%%%%%%%%%%%%%%%%%%%%%%%
\section{Convexity}\label{sec:Convexity}
%%%%%%%%%%%%%%%%%%%%%%%%%%%%%%%%%%%%%%%%%%%%%%%%%%%%%

In order to show the convexity of the 2PI effective action $\Gamma(\phi,\Delta)$, it is convenient to work in terms of the variables $\phi'\equiv\phi$ and $\Delta'\equiv\phi^2+\hbar\Delta$, and the rescaled sources $\mathcal{J}'\equiv\mathcal{J}$ and $\mathcal{K}'\equiv\mathcal{K}/2$. In terms of these variables, the effective action is
\ba
\Gamma(\phi,\Delta)&=&W(\mathcal{J},\mathcal{K})+\mathcal{J}'\phi'+\mathcal{K}'\Delta',
\ea
wherein the dependence of $\mathcal{J}$ and $\mathcal{K}$ on $\phi$ and $\Delta$ has been suppressed. We then have that
\numparts
\ba
\frac{\partial \Gamma(\phi,\Delta)}{\partial \phi'}&=& \frac{\partial\Gamma(\phi,\Delta)}{\partial \phi}\frac{\partial \phi}{\partial\phi'}+\frac{\partial \Gamma(\phi,\Delta)}{\partial \Delta}\frac{\partial \Delta}{\partial \phi'}\nonumber\\&=&\mathcal{J}'+2\mathcal{K}'\phi'-2\mathcal{K}'\phi'=\mathcal{J}',\\
\frac{\partial \Gamma(\phi,\Delta)}{\partial \Delta'}&=&\mathcal{K}',
\ea
\endnumparts
and
\numparts
\ba
\phi'&=&-\frac{\partial W(\mathcal{J},\mathcal{K})}{\partial \mathcal{J}'},\\
\Delta'&=&-\frac{\partial W(\mathcal{J},\mathcal{K})}{\partial \mathcal{K}'}.
\ea
\endnumparts
The variables $\phi'$ and $\Delta'$ are the convex-conjugate variables to $\mathcal{J}$ and $\mathcal{K}$, and they are proportional (up to a sign) to the tangents of the Schwinger function.

\begin{figure}[!t]
	\centering
	\includegraphics[width=60mm]{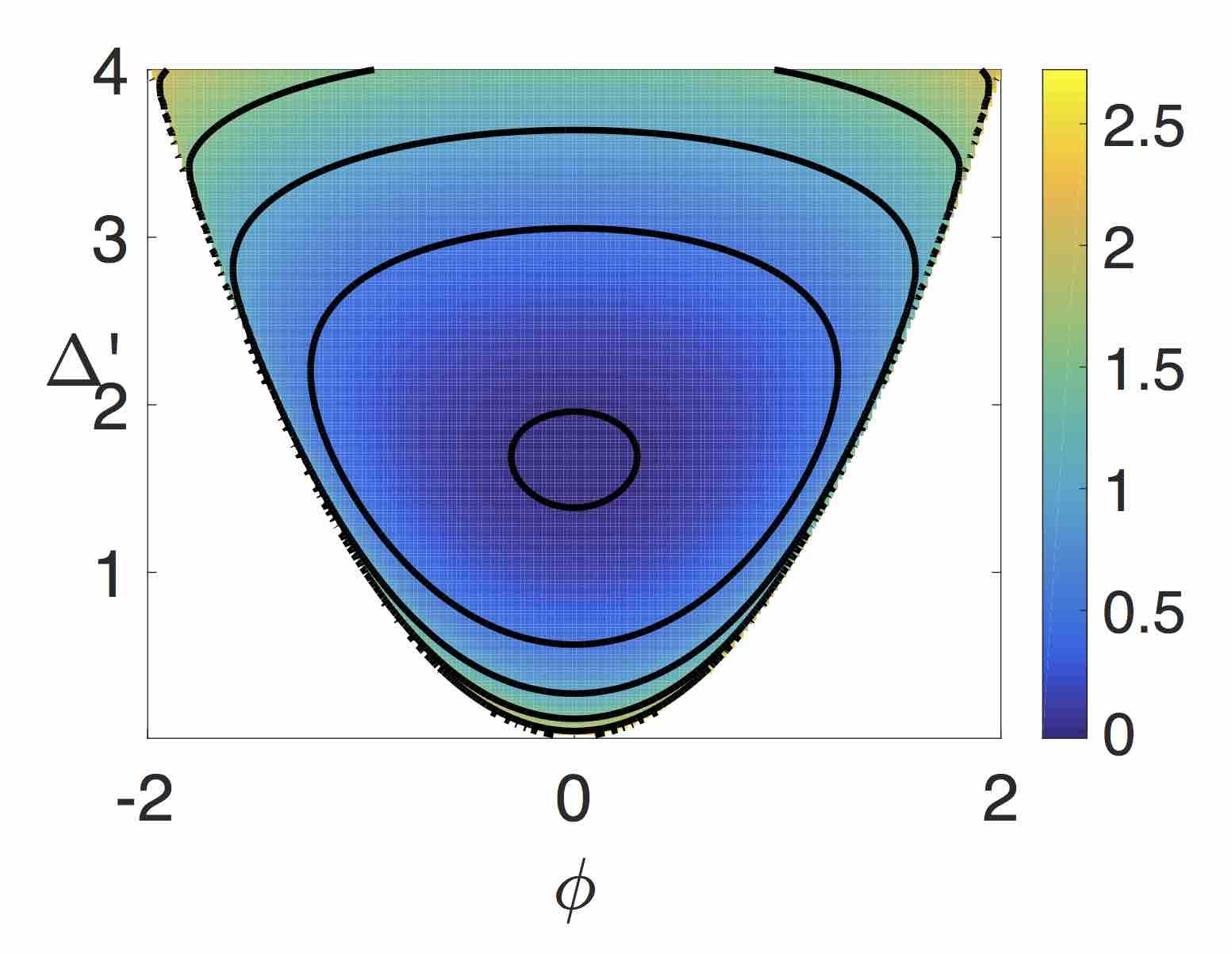}
	\caption{Plot of the effective action as a function of $(\phi,\;\Delta')$ for $m^2=-2$ and $\lambda=6$.\label{fig:convexity}}
\end{figure}

If the effective action is convex with respect to the variables $\phi'$ and $\Delta'$, its Hessian matrix with respect to the variables $\phi'$ and $\Delta'$ must be positive semi-definite (cf.~the 1PI case in \cite{Alexandre:2012ht}). We start by considering the Hessian matrix of $W$ with respect to $\mathcal{J}'$ and $\mathcal{K}'$, given by
\ba
{\rm Hess}(W)(\mathcal{J}',\mathcal{K}')&=&\left(\begin{array}{c c} -\frac{\partial\phi'}{\partial \mathcal{J}'} & -\frac{\partial \phi'}{\partial \mathcal{K}'} \\ -\frac{\partial \Delta'}{\partial \mathcal{J}'} & -\frac{\partial \Delta'}{\partial \mathcal{K}'}\end{array}\right).
\ea
It is the negative of a covariance matrix and therefore negative semi-definite. Specifically, we have that
\numparts
\ba
-\frac{\partial W(\mathcal{J},\mathcal{K})}{\partial \mathcal{J}^{\prime2}}&=&\braket{\Phi^2}-\braket{\Phi}^2=\braket{(\Phi-\braket{\Phi})^2}={\rm cov}(\Phi,\Phi),\\
-\frac{\partial W(\mathcal{J},\mathcal{K})}{\partial \mathcal{K}^{\prime2}}&=&\braket{\Phi^4}-\braket{\Phi^2}^2=\braket{(\Phi^2-\braket{\Phi^2})^2}={\rm cov}(\Phi^2,\Phi^2),\\
-\frac{\partial W(\mathcal{J},\mathcal{K})}{\partial \mathcal{J}'\partial\mathcal{K}'}&=&\braket{\Phi^3}-\braket{\Phi}\braket{\Phi^2}=\braket{(\Phi-\braket{\Phi})(\Phi^2-\braket{\Phi}^2)}={\rm cov}(\Phi,\Phi^2).\nonumber\\
\ea
\endnumparts
The Hessian matrix of $\Gamma$ with respect to the variables $\phi'$ and $\Delta'$ is
\ba
{\rm Hess}(\Gamma)(\phi',\Delta')&=&\left(\begin{array}{c c} \frac{\partial\mathcal{J}'}{\partial \phi'} & \frac{\partial \mathcal{J}'}{\partial \Delta'} \\ \frac{\partial \mathcal{K}'}{\partial \phi'} & \frac{\partial \mathcal{K}'}{\partial\Delta'}\end{array}\right).
\ea
We thus have for the product
\ba
-{\rm Hess}(\Gamma)(\phi',\Delta')\cdot{\rm Hess}(W)(\mathcal{J}',\mathcal{K}')&=&\left(\begin{array}{c c}\frac{{\rm d}\mathcal{J}'}{{\rm d}\mathcal{J}'} & \frac{{\rm d}\mathcal{J}'}{{\rm d}\mathcal{K}'}\\ \frac{{\rm d}\mathcal{K}'}{{\rm d}\mathcal{J}'} & \frac{{\rm d}\mathcal{K}'}{{\rm d}\mathcal{K}'}\end{array}\right)=\mathbb{I},
\ea
since $\mathcal{J}'$ and $\mathcal{K}'$ are independent. The inverse of a negative-definite matrix is a negative-definite matrix, and therefore (ignoring the singular case) the Hessian of $\Gamma$ is positive definite, such that $\Gamma$ is convex with respect to the variables $\phi'$ and $\Delta'$. We remark that it is not, in general, convex with respect to the variables $\phi$ and $\Delta$, as is the case, for example, for a non-convex classical action with $m^2<0$. The situation is illustrated by figures~\ref{fig:GammaJKa} and \ref{fig:convexity}.

%%%%
%%%%

%%%%%%%%%%%%%%%%%%%%%%%%%%%%%%%%%%%%%%%%%%%%%%%%%%%%%
\section{Single saddle point}\label{sec:SingleSaddle}
%%%%%%%%%%%%%%%%%%%%%%%%%%%%%%%%%%%%%%%%%%%%%%%%%%%%%

In order to evaluate the partition function in (\ref{eq:Z}), we can first identify the saddle points $\{\varphi_i\}$ of the classical action in the presence of the sources ${\cal J}(\phi,\Delta)$ and ${\cal K}(\phi,\Delta)$. They are solutions to the stationarity or saddle-point condition
\ba
\label{eq:saddle}
S^{(1)}(\varphi_i)-{\cal J}(\phi,\Delta)-{\cal K}(\phi,\Delta)\varphi_i&=&0,
\ea
where
\ba
S^{(n)}(\varphi_i)\equiv\frac{\partial ^nS(\Phi)}{\partial \Phi^n}\bigg|_{\Phi=\varphi_i},
\ea
and we can introduce the corresponding two-point variables
\ba
\label{eq:Gdef2}
\mathcal{G}_i=\left[G^{-1}(\varphi_i)-\mathcal{K}(\phi,\Delta)\right]^{-1},
\ea
where
\ba
\label{eq:treeG}
G^{-1}(\varphi_i)=S^{(2)}(\varphi_i)=m^2+\frac{\lambda}{2}\varphi_i^2.
\ea
Since the defining equations depend on $\phi$ and $\Delta$ through the sources $\mathcal{J}$ and $\mathcal{K}$, we have that $\varphi_i\equiv\varphi_i(\phi,\Delta)$ and $\mathcal{G}_i\equiv\mathcal{G}_i(\phi,\Delta)$. Moreover, the map $(\phi,\Delta) \to (\varphi_i,\mathcal{G}_i)$ need not be injective, and we will discuss this further in section~\ref{sec:MultiSaddle}.

When the map is injective, and we have a unique saddle point $\varphi$, we can evaluate the integral over $\Phi$ by expanding
\ba
\label{eq:saddlePhi}
\Phi&=&\varphi+\sqrt{\hbar}\hat\Phi,
\ea
giving
\ba\label{eq:S_expanded}
S(\Phi)-{\cal J}\Phi-\frac{1}{2}{\cal K}\Phi^2&=&S(\varphi)-{\cal J}\varphi-\frac{1}{2}{\cal K}\varphi^2+\frac{\hbar}{2!}\mathcal{G}^{-1}\hat\Phi^2\nonumber\\&&\qquad+\frac{\hbar^{3/2}}{3!}S^{(3)}(\varphi)\hat\Phi^3+\frac{\hbar^2}{4!}S^{(4)}(\varphi)\hat\Phi^4,\nonumber\\
\ea
where the absence of the linear term is due to the saddle-point condition in (\ref{eq:saddle}) (cf.~\cite{Ellis:2015xwp}). We may now evaluate $Z({\cal J},{\cal K})$:
\ba
Z({\cal J},{\cal K})&=&\exp\left[ -\frac{1}{\hbar}\left( S(\varphi)-{\cal J}\varphi-\frac{1}{2}{\cal K}\varphi^2 \right) \right]\nonumber\\
	&~&\times \int{\rm d}\hat\Phi\exp\left[ -\frac{\hbar^{1/2}}{3!}\lambda\varphi\hat\Phi^3-\frac{\hbar}{4!}\lambda\hat\Phi^4 \right]\exp\left[ -\frac{1}{2}{\cal G}^{-1}\hat\Phi^2 \right].
\ea
Expanding the exponential and performing the Gaussian integrals, we find
 \ba\nonumber
Z({\cal J},{\cal K})&\approx&\exp\left[ -\frac{1}{\hbar}\left( S(\varphi)-{\cal J}\varphi-\frac{1}{2}{\cal K}\varphi^2 +\frac{\hbar}{2}\ln\mathcal{G}^{-1}G(0)\right) \right]\nonumber\\
		&~&\times\exp\left[ -\frac{\hbar}{8}\lambda{\cal G}^2+\left(\frac{1}{12}+\frac{1}{8}\right)\hbar\lambda^2\varphi^2 {\cal G}^3 \right],\label{eq:Z_JK}
\ea
wherein we have expanded to order $\hbar$ and re-exponentiated the result, for convenience, since we will later take the natural logarithm. We have written $\frac{5}{24}$ as $\frac{1}{12}+\frac{1}{8}$ for illustration, since, in the multi-dimensional field-theory case, this term comes from the sunset plus the dumbell diagrams with the same combinatorical factors. We have absorbed constant factors into the overall normalisation (reset to unity) and introduced the factor of $G(0)$ [see~(\ref{eq:treeG})] to ensure the argument of the logarithm is dimensionless.

We can now use (\ref{eq:effAction}) to find the expression for the effective action
\ba\label{eq:effAct_KJ}
\Gamma(\phi,\Delta)&=&S(\varphi)+\hbar\Gamma_1(\varphi,{\cal G}) +\hbar^2\Gamma_2(\varphi,{\cal G})+\hbar^2\Gamma_{\rm 1PR}(\varphi,{\cal G})\nonumber\\
	&~&+{\cal J}\left(\phi-\varphi\right)+\frac{1}{2}{\cal K}(\phi^2-\varphi^2+\hbar\Delta-\hbar{\cal G}),
\ea
where we have defined
\numparts
\ba
\Gamma_1(\varphi,{\cal G})=\frac{1}{2}\left[ \ln\left( {\cal G}^{-1}G(0)\right)+{\cal K}{\cal G} \right],\\\label{eq:GammaG}
	\phantom{\Gamma_1(\varphi,{\cal G})}=\frac{1}{2}\left[ \ln\left( {\cal G}^{-1}G(0)\right)+G^{-1}{\cal G}-1 \right],\\\label{eq:Gamma_2}
\Gamma_2(\varphi,{\cal G})=\frac{1}{8}\lambda{\cal G}^2-\frac{1}{12}\lambda^2\varphi^2 {\cal G}^3,\\\label{eq:Gamma_1pr}
\Gamma_{\rm 1PR}(\varphi,{\cal G})=-\frac{1}{8}\lambda^2\varphi^2 {\cal G}^3.
\ea
\endnumparts
The subscript ${\rm 1PR}$ labels the one-particle-reducible contribution. 

By virtue of its definition in (\ref{eq:phi}), performing the same expansion around the saddle point, we find
\ba \label{eq:phi_varphi}
\phi&=&\left(1-\frac{\hbar}{2}\lambda{\cal G}^2\right)\varphi,
\ea
which can be inverted to give
\ba
\varphi&=&\left(1+\frac{\hbar}{2}\lambda{\cal G}^2\right)\phi
\ea
Proceeding similarly from (\ref{eq:Delta}), we obtain
\ba\label{eq:DeltaG}
\Delta&=&{\cal G}-\frac{\hbar}{2}\lambda{\cal G}^3+\hbar\lambda^2\varphi^2{\cal G}^4,
\ea
where we have used (\ref{eq:phi_varphi}) to eliminate $\phi$.

Following \cite{Garbrecht:2015cla}, the left-hand side of the expression (\ref{eq:effAct_KJ}) for the effective action may be Taylor expanded about $\varphi$ and ${\cal G}$ to give
\ba\label{eq:GammaTaylor}
\Gamma(\phi,\Delta)&=&\Gamma(\varphi,{\cal G})+\left.\frac{\del\Gamma(\phi,\Delta)}{\del\phi}\right|_{\varphi,{\cal G}}(\phi-\varphi)+\frac{1}{2}\left.\frac{\del^2\Gamma(\phi,\Delta)}{\del\phi^2}\right|_{\varphi,{\cal G}}(\phi-\varphi)^2\nonumber\\
	&~&+\left.\frac{\del\Gamma(\phi,\Delta)}{\del\Delta}\right|_{\varphi,{\cal G}}(\Delta-{\cal G})+\dots ,
\ea
where the subscript ``$\varphi,\mathcal{G}$'' indicates evaluation at $\phi=\varphi$ and $\Delta=\mathcal{G}$. We can also use (\ref{eq:J}) and (\ref{eq:K}) to rewrite the right-hand side of (\ref{eq:effAct_KJ}) as
\ba\label{eq:effAct_KJ_II}
\Gamma(\phi,\Delta)&=&S(\varphi)+\hbar\Gamma_1(\varphi,{\cal G}) +\hbar^2\Gamma_2(\varphi,{\cal G})+\hbar^2\Gamma_{\rm 1PR}(\varphi,{\cal G})\nonumber\\
	&~&+\frac{\del\Gamma(\phi,\Delta)}{\del\phi}\left(\phi-\varphi\right)-\frac{1}{\hbar}\frac{\del\Gamma(\phi,\Delta)}{\del\Delta}\left[(\phi-\varphi)^2-\hbar(\Delta-{\cal G})\right],
\ea
noting that $\frac{\del\Gamma(\phi,\Delta)}{\del\phi}$ and $\frac{\del\Gamma(\phi,\Delta)}{\del\Delta}$ are evaluated at the point $(\phi,\Delta)$. Expanding the first of these further, we have
\ba
\frac{\del\Gamma(\phi,\Delta)}{\del\phi}&=&\left.\frac{\del\Gamma(\phi,\Delta)}{\del\phi}\right|_{\varphi,{\cal G}}+\left.\frac{\del^2\Gamma(\phi,\Delta)}{\del\phi^2}\right|_{\varphi,{\cal G}}(\phi-\varphi)+\dots.
\ea
Equating (\ref{eq:GammaTaylor}) and (\ref{eq:effAct_KJ_II}), we then obtain
\ba\nonumber
\Gamma(\varphi,{\cal G})&=&
	S(\varphi)+\hbar\Gamma_1(\varphi,{\cal G}) +\hbar^2\Gamma_2(\varphi,{\cal G})+\hbar^2\Gamma_{\rm 1PR}(\varphi,{\cal G})\\\label{eq:Gamma_temp1}
		&~&+\frac{1}{2}\left.\frac{\del^2\Gamma(\phi,\Delta)}{\del\phi^2}\right|_{\varphi,{\cal G}}(\phi-\varphi)^2-\frac{1}{\hbar}\frac{\del\Gamma(\phi,\Delta)}{\del\Delta}(\phi-\varphi)^2,
\ea
where the combination
\ba
\label{eq:Kcomb}
\frac{\partial^2\Gamma(\phi,\Delta)}{\partial \phi^2}\bigg|_{\varphi,\mathcal{G}}-\frac{2}{\hbar}\frac{\partial \Gamma(\phi,\Delta)}{\partial \Delta}&=&S^{(2)}(\varphi)-\mathcal{K}(\phi,\Delta)+\mathcal{O}(\hbar)\nonumber\\&=&\mathcal{G}^{-1}+\mathcal{O}(\hbar).
\ea
Making use of (\ref{eq:Gamma_1pr}) and (\ref{eq:phi_varphi}), we can then show that the ${\rm 1PR}$ piece of (\ref{eq:Gamma_temp1}) cancels, leaving
\ba
\Gamma(\varphi,{\cal G})&=&
	S(\varphi)+\hbar\Gamma_1(\varphi,{\cal G}) +\hbar^2\Gamma_2(\varphi,{\cal G}).
\ea
Perhaps unsurprisingly, this is of exactly the same form as the usual expression in terms of $\phi$ and $\Delta$, which we could have found had we expanded the right-hand side of (\ref{eq:effAction}) in terms of $\phi$ rather than $\varphi$, i.e.
\ba\label{eq:CJTaction}
\Gamma(\phi,\Delta)&=&
	S(\phi)+\hbar\Gamma_1(\phi,\Delta) +\hbar^2\Gamma_2(\phi,\Delta).
\ea

%%%%%%%%%%%%%%%%%%%%%%%%%%%%%%%%%%%%%%%%%%%%%%%%%%%%%
\section{Cornwall-Jackiw-Tomboulis 2PI effective action}\label{sec:CJT}
%%%%%%%%%%%%%%%%%%%%%%%%%%%%%%%%%%%%%%%%%%%%%%%%%%%%%

If the system is isolated then we should expect that the physical configuration $(\bar{\varphi},\bar{\mathcal{G}})$ is such that
\ba
\label{eq:vanishing}
\mathcal{J}(\bar{\varphi},\bar{\mathcal{G}})=0\quad {\rm and}\quad \mathcal{K}(\bar{\varphi},\bar{\mathcal{G}})=0,
\ea
i.e.~that for which the sources vanish. We emphasise, as we will see, that $\mathcal{J}(\phi,\Delta)$ and $\mathcal{K}(\phi,\Delta)$ are nevertheless \emph{non-zero} at an arbitrary configuration $(\phi,\Delta)$. The physical configuration then coincides with the extremal point
\numparts
\ba\label{eq:CJTphi}
\left.\frac{\del\Gamma(\phi,\Delta)}{\del\phi}\right|_{\bar{\varphi},\bar{{\cal G}}}&=&0,\\\label{eq:CJTdelta}
\left.\frac{\del\Gamma(\phi,\Delta)}{\del\Delta}\right|_{\bar{\varphi},\bar{\cal G}}&=&0,
\ea
\endnumparts
cf.~figure~\ref{fig:GammaJK}, and we recover the usual interpretation of the Cornwall-Jackiw-Tomboulis 2PI effective action~\cite{Cornwall:1974vz}. We remark that this extremal point is the point at which all $n$PI effective actions coincide (when calculated to all orders), again as illustrated in figure~\ref{fig:GammaJK}. [For a closed or open system, the physical configurations need not correspond to vanishing sources. For instance, at finite temperature, the source $\mathcal{K}(\phi,\Delta)$ is used to encode information about the thermal ensemble (see, e.g., \cite{Calzetta:1986cq, Berges:2004yj, Millington:2012pf}), and we have $\mathcal{K}(\bar{\varphi},\bar{\mathcal{G}})\neq 0$.]

Equation (\ref{eq:CJTphi}), when combined with (\ref{eq:CJTaction}), gives the quantum equation of motion for the physical one-point variable
\ba
\label{eq:hbareom}
\left.\frac{\del S(\phi)}{\del\phi}\right|_{\bar{\varphi},\bar{\cal G}}&=&-\hbar\left.\frac{\del\Gamma_1(\phi,\Delta)}{\del\phi}\right|_{\bar{\varphi},\bar{\cal G}}.
\ea
If the quantum corrections are small, in the sense that the quantum-corrected one-point variable $\bar{\varphi}$ is perturbatively close to the classical one-point variable $\bar{\varphi}_{\rm cl}$, satisfying $S^{(1)}(\bar{\varphi}_{\rm cl})=0$, then we might stop here. However, there are cases where the true quantum configuration of the system is non-perturbatively far away from the classical configuration: an example occurs when metastable states are induced by radiative corrections~\cite{Garbrecht:2015cla, Garbrecht:2015yza}. In such cases, we might hope to improve our perturbation theory by expanding the path integral around the quantum-corrected configuration $\bar{\varphi}$. Having realised, however, that the sources need not vanish for general $\phi$ and $\Delta$, they can be used consistently to drive the saddle point of the partition function towards the physical quantum-corrected configuration. To do so, and closely following \cite{Garbrecht:2015cla} throughout what follows, we simply impose that the saddle point coincides with the physical configuration, and comparing (\ref{eq:hbareom}) with (\ref{eq:saddle}), we obtain the \emph{consistency relation}
\ba
\label{eq:consistency}
{\cal J}(\phi,\Delta)+{\cal K}(\phi,\Delta)\bar{\varphi}&=&-\hbar\left.\frac{\del\Gamma_1(\phi,\Delta)}{\del\phi}\right|_{\bar{\varphi},\bar{\cal G}}
\ea
(to leading order in $\hbar$). Notice that this only constrains \emph{one} linear combination of the sources.

In order to provide an additional constraint on the sources, we can use the Schwinger-Dyson equation, which is obtained from (\ref{eq:GammaG}) and (\ref{eq:CJTaction}), after imposing (\ref{eq:CJTdelta}) and applying (\ref{eq:vanishing}):
\ba\label{eq:dGamma_dDelta_SD}
\bar{{\cal G}}^{-1}&=&G^{-1}(\bar{\varphi})+2\hbar\left.\frac{\del\Gamma_2(\phi,\Delta)}{\del\Delta}\right|_{\bar{\varphi},\bar{\cal G}}.
\ea
Comparing this with the definition of $\bar{\mathcal{G}}^{-1}$ in (\ref{eq:Gdef2}), we therefore have that
\ba\label{eq:K_Gamma2}
{\cal K}(\phi,\Delta)&=&-2\hbar\left.\frac{\del\Gamma_2(\phi,\Delta)}{\del\Delta}\right|_{\bar{\varphi},\bar{{\cal G}}}.
\ea
Inserting this expression for ${\cal K}$ into the consistency relation in (\ref{eq:consistency}), we can fix
\ba
\label{eq:J_Gamma2}
{\cal J}(\phi,\Delta)&=&-\hbar\left.\frac{\del\Gamma_1(\phi,\Delta)}{\del\phi}\right|_{\bar{\varphi},\bar{{\cal G}}}+2\hbar\left.\frac{\del\Gamma_2(\phi,\Delta)}{\del\Delta}\right|_{\bar{\varphi},\bar{\cal G}}\bar{\varphi}.
\ea
We see that both sources are order $\hbar$ and that their role in ensuring that the saddle point coincides with the physical configuration is to put the loop corrections into the exponent of the partition function.

In order to show that the above procedure is self-consistent, we need to confirm that the expressions for the sources in (\ref{eq:K_Gamma2}) and (\ref{eq:J_Gamma2}) are consistent with (\ref{eq:vanishing}). In order to do so, we first note that, since the sources are order $\hbar$ and the saddle point is unique, $\phi$ and $\bar{\varphi}$, and $\Delta$ and $\bar{\mathcal{G}}$ differ by terms of order $\hbar$.

Starting with expression (\ref{eq:J}), we can therefore expand in $\phi-\bar{\varphi}$ and $\Delta-\bar{\mathcal{G}}$ to give
\ba\nonumber
{\cal J}(\phi,\Delta) +{\cal K}(\phi,\Delta)\phi &=& \left.\frac{\del\Gamma(\phi,\Delta)}{\del\phi}\right|_{\bar{\varphi},\bar{\cal G}}+\left.\frac{\del^2\Gamma(\phi,\Delta)}{\del\phi^2}\right|_{\bar{\varphi},\bar{\cal G}}(\phi-\bar{\varphi})
	\nonumber\\&&+\left.\frac{\del^2\Gamma(\phi,\Delta)}{\del\phi\del\Delta}\right|_{\bar{\varphi},\bar{\cal G}}(\Delta-\bar{\cal G}).
\ea
The first term on the right-hand side gives
\ba
\left.\frac{\del\Gamma(\phi,\Delta)}{\del\phi}\right|_{\bar{\varphi},\bar{\cal G}}=\mathcal{J}(\bar{\varphi},\bar{\mathcal{G}})+\mathcal{K}(\bar{\varphi},\bar{\mathcal{G}})\bar{\varphi}.
\ea
From (\ref{eq:GammaG}), we have that
\ba
\Gamma_1(\phi,\Delta)=\frac{1}{2}\ln[\Delta^{-1}G(0)]+\frac{1}{2}[G^{-1}(\phi)\Delta-1],
\ea
and so
\ba
\frac{\del \Gamma_1(\phi,\Delta)}{\del\phi}&=&\frac{1}{2}\frac{\del G^{-1}(\phi)}{\del\phi}\Delta
=\frac{1}{2}\lambda\phi\Delta.
\ea
Using this result along with (\ref{eq:phi_varphi}) and (\ref{eq:Kcomb}), and noting from (\ref{eq:K}) and (\ref{eq:DeltaG}) that $\frac{\del\Gamma}{\del\Delta}\sim\hbar{\cal K}\sim\hbar^2$ and $\Delta-{\cal G}\sim\hbar$, we obtain
\ba\nonumber
{\cal J}(\phi,\Delta) +{\cal K}(\phi,\Delta)\phi
	&=&\mathcal{J}(\bar{\varphi},\bar{\mathcal{G}})+\mathcal{K}(\bar{\varphi},\bar{\mathcal{G}})\bar{\varphi}+\frac{\partial^2\Gamma(\phi,\Delta)}{\partial\phi^2}\bigg|_{\bar{\varphi},\bar{\mathcal{G}}}(\phi-\bar{\varphi})\nonumber\\&=&\mathcal{J}(\bar{\varphi},\bar{\mathcal{G}})+\mathcal{K}(\bar{\varphi},\bar{\mathcal{G}})\bar{\varphi}+\mathcal{G}^{-1}\left(-\frac{1}{2}\hbar\lambda\bar{\varphi}\bar{\cal G}^2\right)\nonumber\\&=&\mathcal{J}(\bar{\varphi},\bar{\mathcal{G}})+\mathcal{K}(\bar{\varphi},\bar{\mathcal{G}})\bar{\varphi}-\hbar\frac{\partial \Gamma_1(\phi,\Delta)}{\partial \phi}\bigg|_{\bar{\varphi},\bar{\mathcal{G}}}.
\ea
Since ${\cal K}\sim\hbar$, we can replace $\mathcal{K}(\phi,\Delta)\phi\to\mathcal{K}(\phi,\Delta)\bar{\varphi}$ at the order we are working, and we have
\ba
{\cal J}(\phi,\Delta) +{\cal K}(\phi,\Delta)\bar{\varphi}&=&\mathcal{J}(\bar{\varphi},\bar{\mathcal{G}})+\mathcal{K}(\bar{\varphi},\bar{\mathcal{G}})\bar{\varphi}-\hbar\left.\frac{\del \Gamma_1(\phi,\Delta)}{\del\phi}\right|_{\bar{\varphi},\bar{\cal G}}.
\ea
Comparing this with the consistency relation (\ref{eq:consistency}), it follows that
\ba\label{eq:Jconsistency}
{\cal J}(\bar{\varphi},\bar{\cal G}) +{\cal K}(\bar{\varphi},\bar{\cal G})\bar{\varphi}&=&0,
\ea
as required.

In order to show that ${\cal K}(\bar{\varphi},\bar{\cal G})=0$, we proceed similarly, expanding
\ba
{\cal K}(\phi,\Delta)&=&{\cal K}(\bar{\varphi},\bar{\cal G})+\left. \frac{\del{\cal K}(\phi,\Delta)}{\del\phi} \right|_{\bar{\varphi},\bar{\cal G}}(\phi-\bar{\varphi})+\left. \frac{\del{\cal K}(\phi,\Delta)}{\del\Delta} \right|_{\bar{\varphi},\bar{\cal G}}(\Delta-\bar{\cal G}).
\ea
Making use of (\ref{eq:K}), this can be written in terms of derivatives of the effective action as follows:
\ba
\label{eq:Kmidstep}
{\cal K}(\phi,\Delta)&=&{\cal K}(\bar{\varphi},\bar{\cal G})+\frac{2}{\hbar}\left. \frac{\del^2\Gamma(\phi,\Delta)}{\del\phi\del\Delta} \right|_{\bar{\varphi},\bar{\cal G}}(\phi-\bar{\varphi})\nonumber\\&&+\frac{2}{\hbar}\left. \frac{\del^2\Gamma(\phi,\Delta)}{\del\Delta\del\Delta} \right|_{\bar{\varphi},\bar{\cal G}}(\Delta-\bar{\cal G}).
\ea
Since $\phi$ and $\Delta$ are independent, we have that $\frac{\del S(\phi)}{\del\Delta}=0$, and the leading derivative terms arise from $\Gamma_1(\phi,\Delta)$:
\ba
{\cal K}(\phi,\Delta)&=&{\cal K}(\bar{\varphi},\bar{\cal G})+2\left. \frac{\del^2\Gamma_1(\phi,\Delta)}{\del\phi\del\Delta} \right|_{\bar{\varphi},\bar{\cal G}}(\phi-\bar{\varphi})\nonumber\\&&+2\left. \frac{\del^2\Gamma_1(\phi,\Delta)}{\del\Delta\del\Delta} \right|_{\bar{\varphi},\bar{\cal G}}(\Delta-\bar{\cal G}).
\ea
Now, from (\ref{eq:Gdef2}), (\ref{eq:GammaG}) and (\ref{eq:Gamma_2}), we have
\numparts
\ba
\frac{\del\Gamma_1(\phi,\Delta)}{\del\Delta}&=&\frac{1}{2}G^{-1}(\phi)-\frac{1}{2}\Delta^{-1},\\
\frac{\del^2\Gamma_1(\phi,\Delta)}{\del\phi\del\Delta}&=&\frac{1}{2}\lambda\phi,\\
\frac{\del^2\Gamma_1(\phi,\Delta)}{\del\Delta\del\Delta}&=&\frac{1}{2}\Delta^{-2},\\
\frac{\del\Gamma_2(\phi,\Delta)}{\del\Delta}&=&\frac{1}{4}\lambda\Delta-\frac{1}{4}\lambda^2\phi^2\Delta^2.
\ea
\endnumparts
Combining these results with (\ref{eq:phi_varphi}) and (\ref{eq:DeltaG}), we can then show that
\ba
&&2\left. \frac{\del^2\Gamma_1(\phi,\Delta)}{\del\phi\del\Delta} \right|_{\bar{\varphi},\bar{\cal G}}(\phi-\bar{\varphi})+2\left. \frac{\del^2\Gamma_1(\phi,\Delta)}{\del\Delta\del\Delta} \right|_{\bar{\varphi},\bar{\cal G}}(\Delta-\bar{\cal G})
	\nonumber\\&&=-\frac{\hbar}{2}\lambda\bar{\cal G}+\frac{\hbar}{2}\lambda^2\bar{\varphi}^2\bar{\cal G}^2=-2\hbar\left.\frac{\del\Gamma_2(\phi,\Delta)}{\del\Delta}\right|_{\bar{\varphi},\bar{\cal G}}.
\ea
Hence, returning to (\ref{eq:Kmidstep}), we have that
\ba\label{eq:K_KGamma2}
{\cal K}(\phi,\Delta)&=&{\cal K}(\bar{\varphi},\bar{\cal G})-2\hbar\left.\frac{\del\Gamma_2(\phi,\Delta)}{\del\Delta}\right|_{\bar{\varphi},\bar{\cal G}},
\ea
and comparing this with (\ref{eq:K_Gamma2}), it immediately follows that
\ba\label{eq:Kconsistency}
{\cal K}(\bar{\varphi},\bar{\cal G})&=&0,
\ea
again as required. The two relations (\ref{eq:Jconsistency}) and (\ref{eq:Kconsistency}) then prove that, to leading order in $\hbar$, the CJT equations (\ref{eq:CJTphi}) and (\ref{eq:CJTdelta}) are satisfied, if we constrain the external sources such that $\bar{\varphi}$ and $\bar{\cal G}$ are the extrema of the quantum effective action, and once we recall (\ref{eq:J}) and (\ref{eq:K}), as first pointed out in \cite{Garbrecht:2015cla}.

Before concluding this section, we remark that we need not have used the Schwinger-Dyson equation to constrain the source $\mathcal{K}(\phi,\Delta)$. In the case of global symmetries, for instance, we might instead use the Ward identities directly to constrain this source, as was discussed in detail in \cite{Garbrecht:2015cla} (cf.~the methodology of \cite{Pilaftsis:2013xna}). Further study of this use of the sources in zero dimensions will be presented elsewhere.

%%%%%%%%%%%%%%%%%%%%%%%%%%%%%%%%%%%%%%%%%%%%%%%%%%%%%
\section{Multiple saddle points and the Maxwell construction}\label{sec:MultiSaddle}
%%%%%%%%%%%%%%%%%%%%%%%%%%%%%%%%%%%%%%%%%%%%%%%%%%%%%

%%%%
\begin{figure}[!t]
	\centering
	\includegraphics[width=60mm]{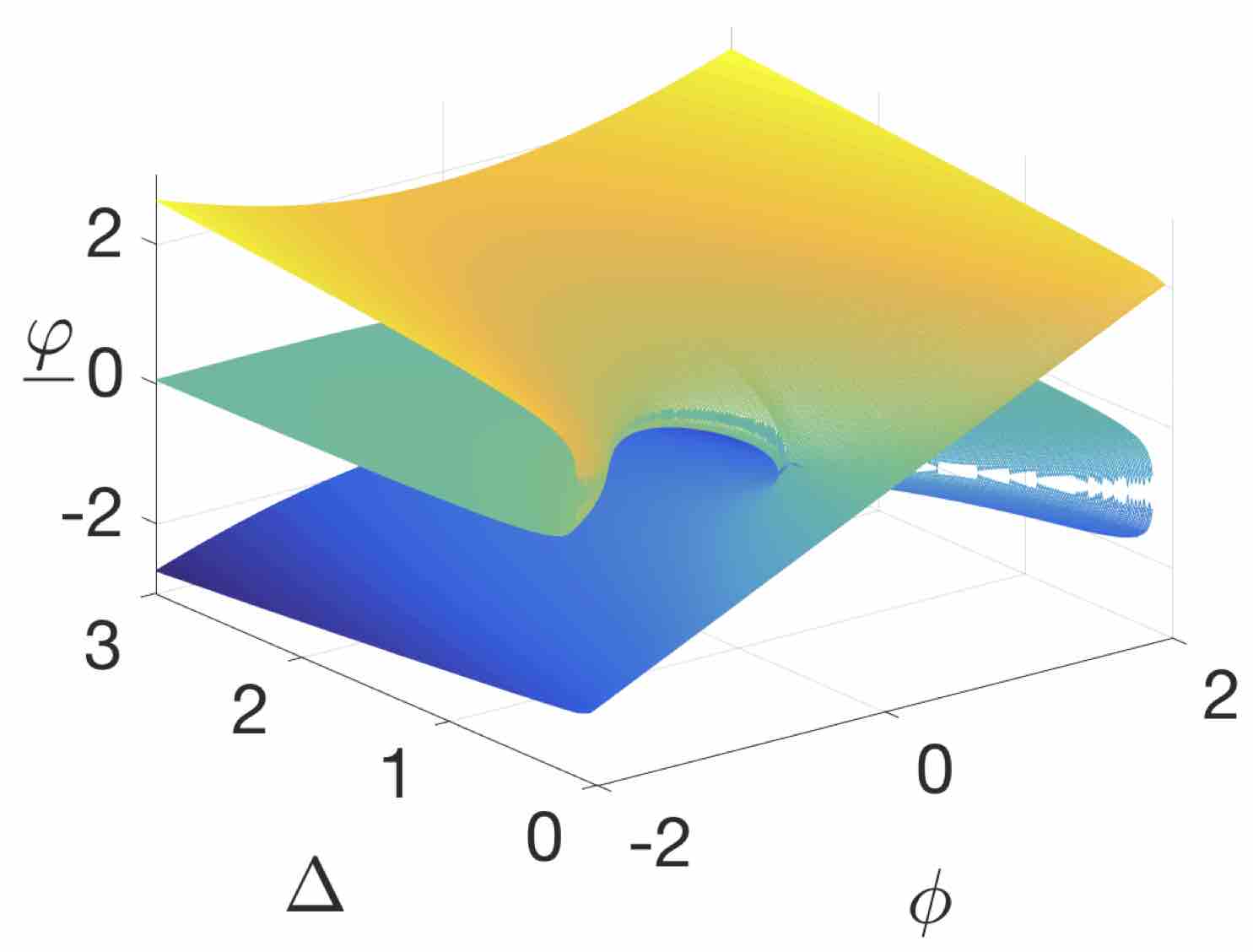}
	\caption{Plot showing the set of saddles $\{\varphi_i\}$ as a function of $\phi$ and $\Delta$ for $m^2=-1$ and $\lambda=6$.\label{fig:all_the_saddles}}
\end{figure}
%%%%

We now turn our attention to the case when the potential has multiple minima, such that there are multiple relevant saddle points $\{\varphi_i\}$. In fact, even for a convex classical potential $V(\Phi)$ ($\equiv S(\Phi)$ in our zero-dimensional setting), we can always choose $\mathcal{K}(\phi,\Delta)$ such that there is a non-convex region. That is, given $V''(\Phi)>0$ over some interval of $\Phi$, we can choose $\mathcal{K}(\phi,\Delta)>V''(\Phi)$ such that $V''_{\mathcal{K}}(\Phi)\equiv V''(\Phi)-\mathcal{K}(\phi,\Delta)<0$ over the same interval. Notice that the number of saddles need not be fixed as a function $\phi$, and this is illustrated explicitly in figure~\ref{fig:all_the_saddles} for $m^2=-1$ and $\lambda=6$.

To evaluate the integral (\ref{eq:Z}), we expand about each of the saddles by writing
\ba
\label{eq:multiSaddlePhi}
\Phi_i&=&\varphi_i+\sqrt{\hbar}\hat\Phi_i.
\ea
Summing up the result from each saddle, we can approximate
\ba
Z(\mathcal{J},\mathcal{K})\approx\sum_iZ_i(\mathcal{J},\mathcal{K}).
\ea
Equation (\ref{eq:S_expanded}) is then modified simply to an expression in the region of each saddle by $\varphi\to\varphi_i$ and $\hat{\Phi}\to\hat{\Phi}_i$. If we track this through then the equivalent of (\ref{eq:Z_JK}) becomes
 \ba\nonumber
Z({\cal J},{\cal K})&\approx&\sum_i\exp\left[ -\frac{1}{\hbar}\left( S(\varphi_i)-{\cal J}(\phi,\Delta)\varphi_i-\frac{1}{2}{\cal K}(\phi,\Delta)\varphi_i^2 \right.\right.\\
		&~&\left. \left. +\frac{\hbar}{2}\ln\mathcal{G}_i^{-1}G(0)+\frac{\hbar^2}{8}\lambda{\cal G}_i^2-\left(\frac{1}{12}+\frac{1}{8}\right)\hbar^2\lambda^2\varphi_i^2 {\cal G}_i^3\right) \right].\label{eq:Z_JK_multiSaddle}
\ea
In the remainder of this section, we drop the arguments on $\mathcal{J}$ and $\mathcal{K}$ for convenience.

\begin{figure}[t]
\centering
\includegraphics[scale=0.55]{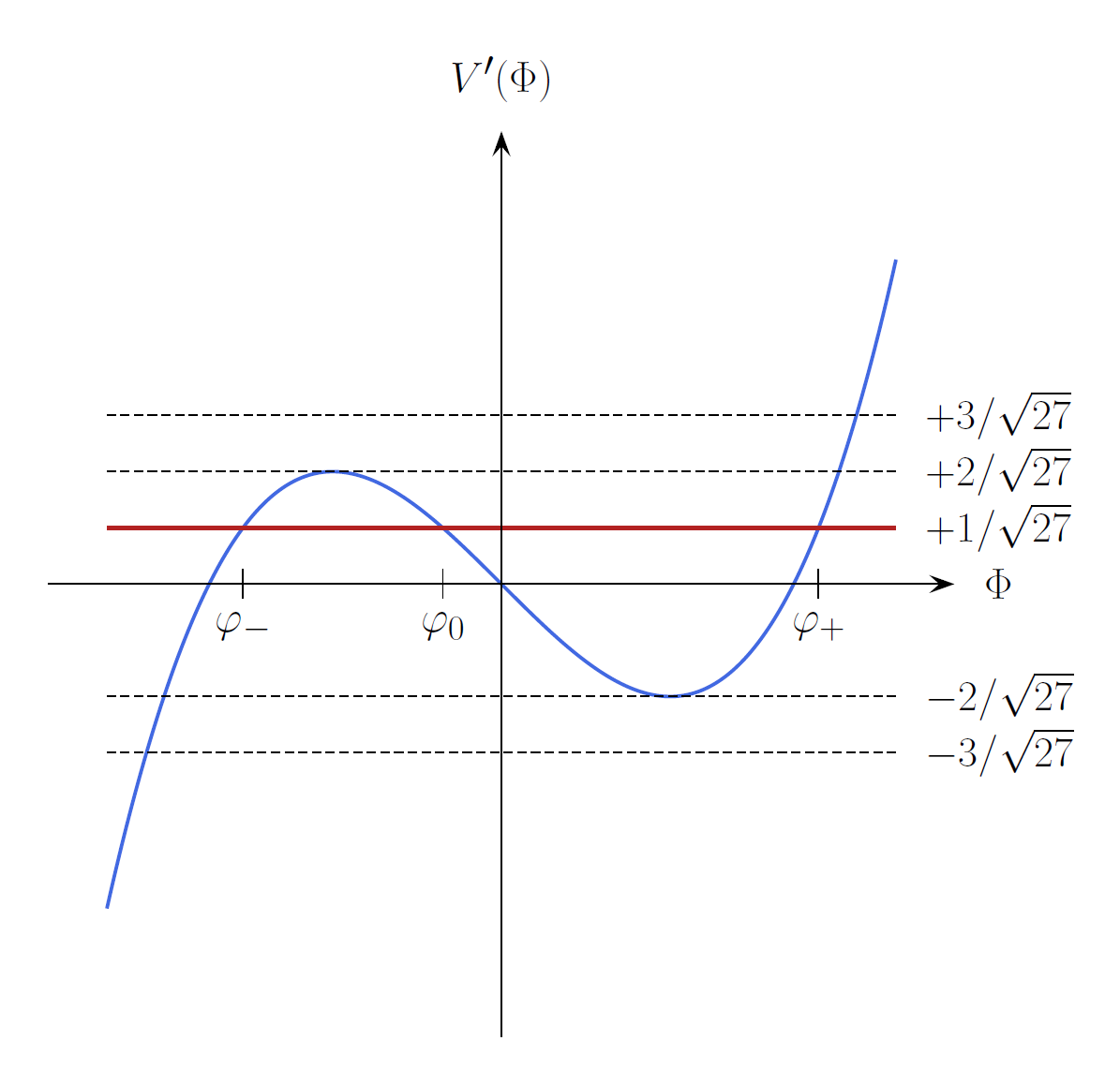}
\caption{\label{fig:ssbMaxwell} Sketch of $V'(\Phi)=-\Phi+\Phi^3$ (i.e.~$m^2=-1$ and $\lambda=6$; blue line) along with a range of values of $\mathcal{J}(\phi)$ for $\mathcal{K}(\phi,\Delta)=0$. The solutions of $V'(\varphi_i)=\mathcal{J}(\phi)$ indicated on the horizontal axis correspond to $\mathcal{J}(\phi)=1/\sqrt{27}$ (red line). For $|\mathcal{J}(\phi)|>2/\sqrt{27}$, there is only one extremum, a minimum. At $\mathcal{J}(\phi)=\pm 2/\sqrt{27}$, we have one minimum and one inflection point.}
\end{figure}

\begin{figure}[t]
\centering
\includegraphics[scale=0.55]{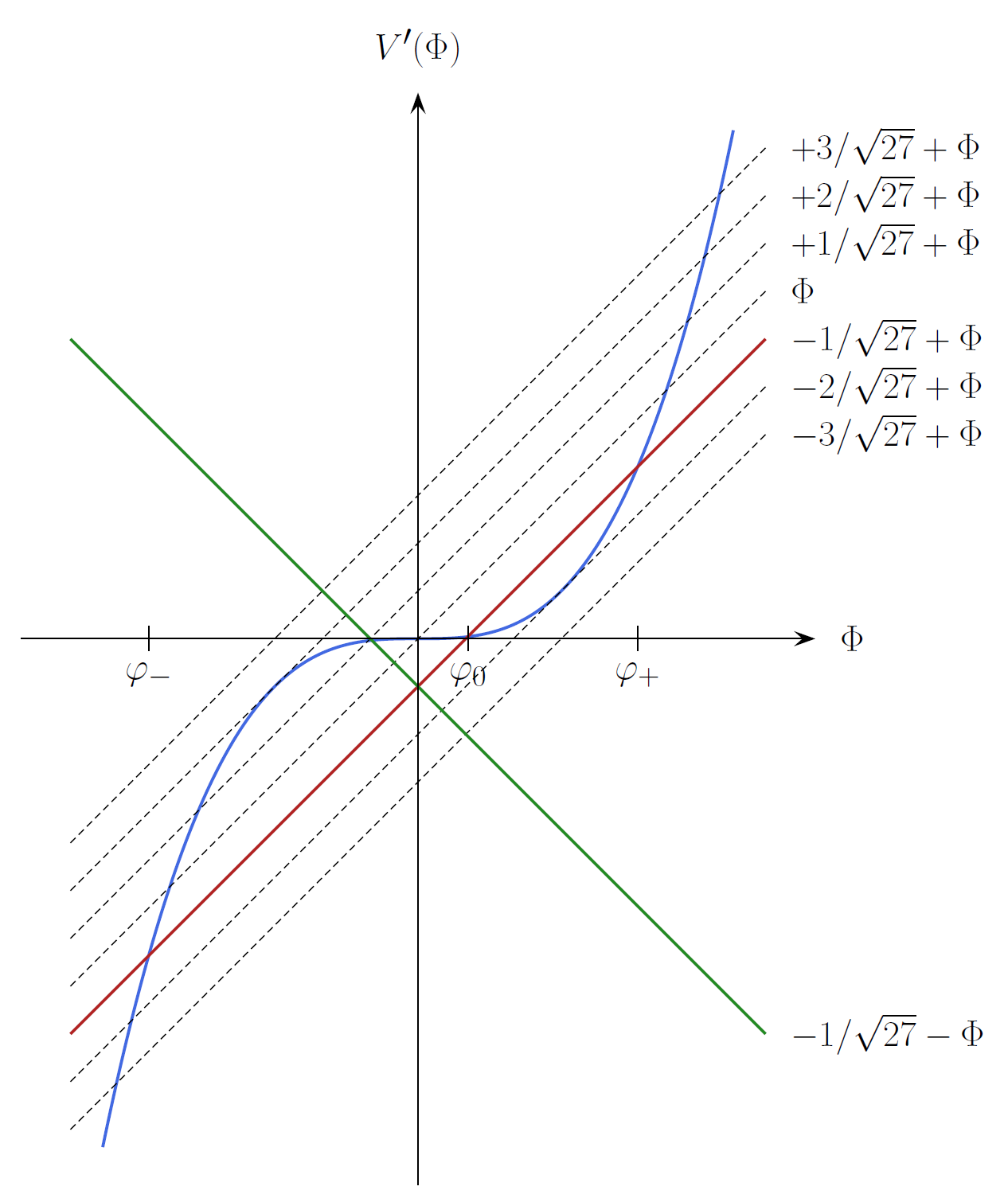}
\caption{\label{fig:nossbMaxwell} Sketch of $V'(\Phi)=\Phi^3$ (i.e.~$m^2=0$ and $\lambda=6$; blue line) alongside a range of $\mathcal{J}(\phi)+\mathcal{K}(\phi,\Delta)\Phi$ for $|\mathcal{K}(\phi,\Delta)|=1$. The solutions of $V'(\varphi_i)=\mathcal{J}(\phi)+\mathcal{K}(\phi,\Delta)\varphi_i$ indicated on the horizontal axis correspond to $\mathcal{J}(\phi)=-1/\sqrt{27}$ and $\mathcal{K}(\phi,\Delta)=1$ (red line). The variation in number and type of extrema with $\mathcal{J}(\phi,\Delta)$ and $\mathcal{K}(\phi,\Delta)$ is again visible. For $\mathcal{K}(\phi,\Delta)\leq 0$ (green line), we have a single saddle point.}
\end{figure}

Let us now suppose that there are two minima at $\varphi_-$ and $\varphi_+$, with $\varphi_-<\varphi_+$. It follows  that (to zeroth order in $\hbar$)
\numparts
\ba
\label{eq:onepointMax}
\phi&\approx&\frac{\varphi_-Z_-+\varphi_+Z_+}{Z_-+Z_+},\\
\Delta'&\approx&\frac{\left(\mathcal{G}_-+\varphi_-^2/\hbar\right)Z_-+\left(\mathcal{G}_++\varphi_+^2/\hbar\right)Z_+}{Z_-+Z_+},
\ea
\endnumparts
from which we find
\ba
\frac{\phi-\varphi_-}{\varphi_+-\phi} =  \frac{Z_+}{Z_-}=\frac{\Delta'-\left(\mathcal{G}_-+\varphi_-^2/\hbar\right)}{\left(\mathcal{G}_++\varphi_+^2/\hbar\right)-\Delta'}.
\ea
(The contribution of the central saddle is negligible, as shown in figure~\ref{fig:saddle_contribution}, see appendix~\ref{app:A}.) We therefore have that (up to and including terms at order $\hbar$)
\ba
-\hbar \ln\left[\frac{\phi-\varphi_-}{\varphi_+-\phi}\right]&=&S_+-S_--\mathcal{J}(\varphi_+-\varphi_-)-\frac{1}{2}\mathcal{K}(\varphi_+^2-\varphi_-^2)\nonumber\\&&+\frac{\hbar}{2}\ln\mathcal{G}^{-1}_+\mathcal{G}_-.
\ea
Rearranging for $\mathcal{J}$, we obtain
\ba
\mathcal{J}&=&\frac{S_+-S_-}{\varphi_+-\varphi_-}-\frac{1}{2}\,\mathcal{K}(\varphi_++\varphi_-)\nonumber\\&&+\frac{\hbar}{\varphi_+-\varphi_-}\left\{\ln\left[\frac{\phi-\varphi_-}{\varphi_+-\phi}\right]+\frac{1}{2}\ln\left[\frac{S^{(2)}_+-\mathcal{K}}{S^{(2)}_--\mathcal{K}}\right]\right\}.
\ea
For $\varphi_-<\phi<\varphi_+$, the argument of the logarithm remains positive. However, we see that we hit branch points at $\phi=\varphi_{\pm}$. This marks the breakdown of the approximation, beyond which we have only one saddle-point configuration. This is illustrated graphically in figures~\ref{fig:ssbMaxwell} and \ref{fig:nossbMaxwell}. We also note that for $\varphi_-<0<\varphi_+$ and fixed $\mathcal{K}$, $\mathcal{J}$ grows approximately linearly with $\phi$ about $\phi=0$.

%%%%
\begin{figure}[!t]
	\centering
	\includegraphics[width=70mm]{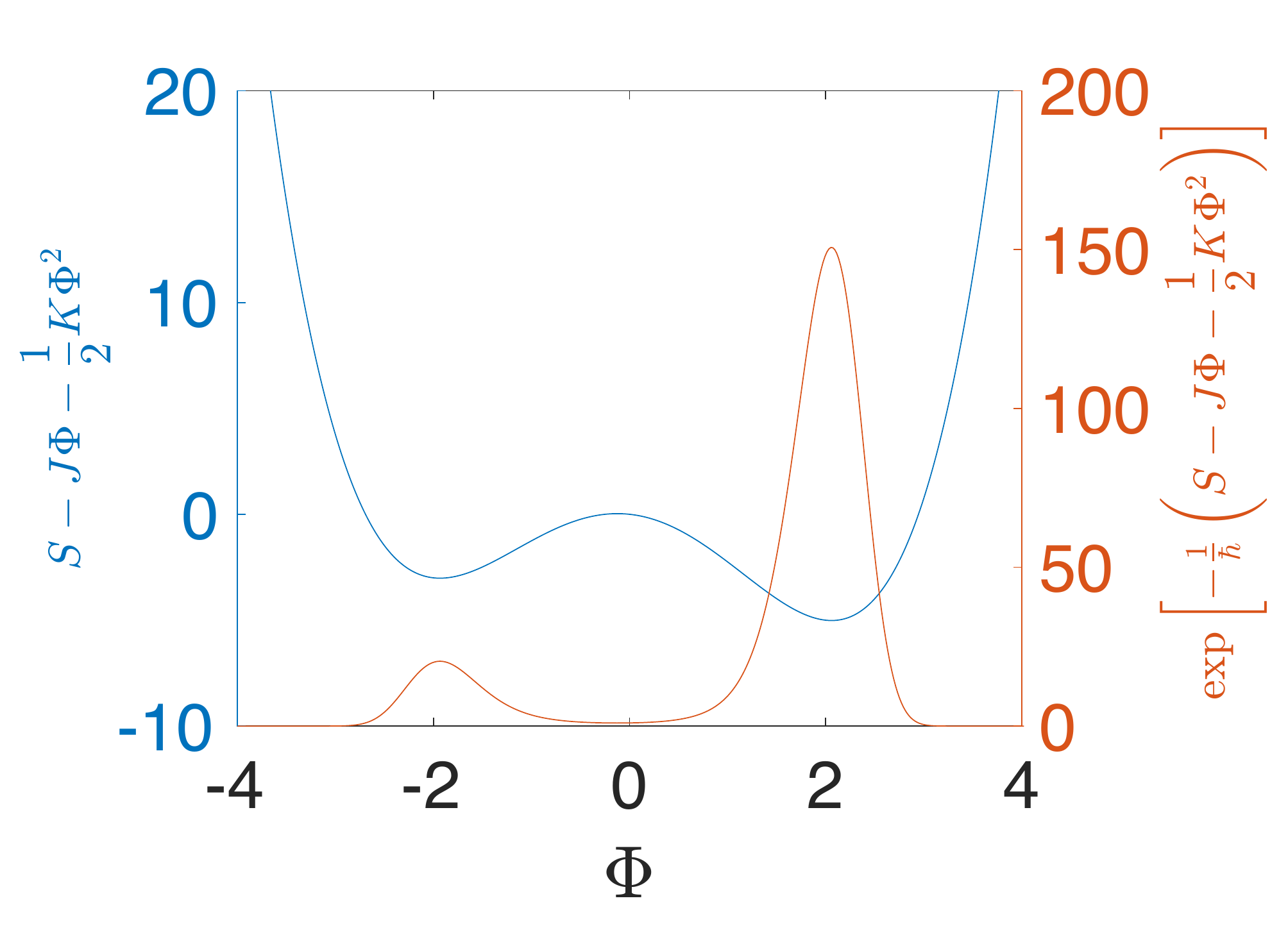}
	\caption{Plot showing the exponent of the exponential and the exponential appearing in the integrand in (\ref{eq:Z}) for $m^2=-1$ and $\lambda=6$. The local maximum corresponds to the largest exponent, and its contribution is therefore exponentially suppressed relative to those of the two minima.\label{fig:saddle_contribution}}
\end{figure}
%%%%

Returning to the effective action, the exponents of $Z_{\pm}$ become
\numparts
\ba
&&S_+-\mathcal{J}\varphi_+-\frac{1}{2}\mathcal{K}\varphi_+^2+\frac{\hbar}{2}\ln(S^{(2)}_+-\mathcal{K})G(0)\nonumber\\&&=\frac{\varphi_+S_--\varphi_-S_+}{\varphi_+-\varphi_-}+\frac{1}{2}\mathcal{K}\varphi_+\varphi_--\frac{\hbar}{2}\ln\frac{\left[\left(S_+^{(2)}-\mathcal{K}\right)G(0)\right]^{\frac{\varphi_-}{\varphi_+-\varphi_-}}}{\left[\left(S_-^{(2)}-\mathcal{K}\right)G(0)\right]^{\frac{\varphi_+}{\varphi_+-\varphi_-}}}\nonumber\\&&-\hbar\ln\left[\frac{\phi-\varphi_-}{\varphi_+-\phi}\right]^{\frac{\varphi_+}{\varphi_+-\varphi_-}},\\
&&S_--\mathcal{J}\varphi_--\frac{1}{2}\mathcal{K}\varphi_-^2+\frac{\hbar}{2}\ln(S^{(2)}_--\mathcal{K})G(0)\nonumber\\&&=\frac{\varphi_+S_--\varphi_-S_+}{\varphi_+-\varphi_-}+\frac{1}{2}\mathcal{K}\varphi_+\varphi_--\frac{\hbar}{2}\ln\frac{\left[\left(S_+^{(2)}-\mathcal{K}\right)G(0)\right]^{\frac{\varphi_-}{\varphi_+-\varphi_-}}}{\left[\left(S_-^{(2)}-\mathcal{K}\right)G(0)\right]^{\frac{\varphi_+}{\varphi_+-\varphi_-}}}\nonumber\\&&-\hbar\ln\left[\frac{\phi-\varphi_-}{\varphi_+-\phi}\right]^{\frac{\varphi_-}{\varphi_+-\varphi_-}},
\ea
\endnumparts
such that
\ba
\Gamma(\phi,\Delta)& = &\frac{(\varphi_+-\phi)\Gamma_-+(\phi-\varphi_-)\Gamma_+}{\varphi_+-\varphi_-}-\frac{1}{2}\mathcal{K}(\varphi_+-\phi)(\phi-\varphi_-)\nonumber\\&&-\hbar\ln\left[\left(\frac{\phi-\varphi_-}{\varphi_+-\phi}\right)^{\frac{\varphi_+-\phi}{\varphi_+-\varphi_-}}+\left(\frac{\varphi_+-\phi}{\phi-\varphi_-}\right)^{\frac{\phi-\varphi_-}{\varphi_+-\varphi_-}}\right]+\frac{\hbar}{2}\mathcal{K}\Delta,
\ea
where
\ba
\Gamma_{\pm}\equiv S_{\pm}+\frac{\hbar}{2}\ln \left[\left(S^{(2)}_{\pm}-\mathcal{K}\right)G(0)\right]
\ea
are the effective actions around each saddle.

%%%%
%%%%

%%%%
%%%%

We recall that $\varphi_{\pm}\equiv\varphi_{\pm}(\phi,\Delta)$. However, to a fixed order in $\hbar$, we can make the dependence on $\phi$ explicit by writing $\varphi_{\pm}(\phi,\Delta)=\tilde{\varphi}_{\pm}+\hbar \delta\varphi_{\pm}(\phi,\Delta)$, so long as $\phi$ and $\Delta$ are such that the logarithms remain small. The equations of motion for the one-point functions $\varphi_{\pm}$ are
\ba
S^{(1)}_{\pm}-\mathcal{K}\varphi_{\pm}&=&\frac{S_+-S_-}{\varphi_+-\varphi_-}-\frac{1}{2}\mathcal{K}\left(\varphi_++\varphi_-\right)\nonumber\\&&+\frac{\hbar}{\varphi_+-\varphi_-}\left[\ln\left(\frac{\phi-\varphi_-}{\varphi_+-\phi}\right)+\frac{1}{2}\ln\frac{S_+^{(2)}-\mathcal{K}}{S_-^{(2)}-\mathcal{K}}\right].\nonumber\\
\ea
Equating terms at zeroth order in $\hbar$, we have
\ba
\tilde{S}^{(1)}_{\pm}-\mathcal{K}\tilde{\varphi}_{\pm}=\frac{\tilde{S}_+-\tilde{S}_-}{\tilde{\varphi}_+-\tilde{\varphi}_-}-\frac{1}{2}\mathcal{K}\left(\tilde{\varphi}_++\tilde{\varphi}_-\right),
\ea
where $\tilde{S}_{\pm}\equiv S(\tilde{\varphi}_{\pm})$.  Equating terms at order $\hbar$, we have
\ba
\left(\tilde{S}_{\pm}^{(1)}-\mathcal{K}\right)\delta \varphi_{\pm}=\frac{1}{\tilde{\varphi}_+-\tilde{\varphi}_-}\left[\ln\left(\frac{\phi-\tilde{\varphi}_-}{\tilde{\varphi}_+-\phi}\right)+\frac{1}{2}\ln\frac{\tilde{S}_+^{(2)}-\mathcal{K}}{\tilde{S}_-^{(2)}-\mathcal{K}}\right],
\ea
wherein all other order-$\hbar$ corrections have cancelled. Proceeding in the same way for the effective action, we find
\ba
\Gamma(\phi,\Delta)& = &\frac{(\tilde{\varphi}_+-\phi)\tilde{\Gamma}_-+(\phi-\tilde{\varphi}_-)\tilde{\Gamma}_+}{\tilde{\varphi}_+-\tilde{\varphi}_-}-\frac{1}{2}\mathcal{K}(\tilde{\varphi}_+-\phi)(\phi-\tilde{\varphi}_-)\nonumber\\&&-\hbar\ln\left[\left(\frac{\phi-\tilde{\varphi}_-}{\tilde{\varphi}_+-\phi}\right)^{\frac{\tilde{\varphi}_+-\phi}{\tilde{\varphi}_+-\tilde{\varphi}_-}}+\left(\frac{\tilde{\varphi}_+-\phi}{\phi-\tilde{\varphi}_-}\right)^{\frac{\phi-\tilde{\varphi}_-}{\tilde{\varphi}_+-\tilde{\varphi}_-}}\right]+\frac{\hbar}{2}\mathcal{K}\Delta,
\ea
where
\ba
\tilde{\Gamma}_{\pm}\equiv \tilde{S}_{\pm}+\frac{\hbar}{2}\ln\left[\left(\tilde{S}_{\pm}^{(2)}-\mathcal{K}\right)G(0)\right].
\ea
In the limit $\mathcal{K}\to 0$, we recover the 1PI result, presented in \cite{Alexandre:2012hn},
\ba
\Gamma(\phi)& = &\frac{(\tilde{\varphi}_+-\phi)\tilde{\Gamma}_-+(\phi-\tilde{\varphi}_-)\tilde{\Gamma}_+}{\tilde{\varphi}_+-\tilde{\varphi}_-}\nonumber\\&&-\hbar\ln\left[\left(\frac{\phi-\tilde{\varphi}_-}{\tilde{\varphi}_+-\phi}\right)^{\frac{\tilde{\varphi}_+-\phi}{\tilde{\varphi}_+-\tilde{\varphi}_-}}+\left(\frac{\tilde{\varphi}_+-\phi}{\phi-\tilde{\varphi}_-}\right)^{\frac{\phi-\tilde{\varphi}_-}{\tilde{\varphi}_+-\tilde{\varphi}_-}}\right],\nonumber\\
\ea
which shows that, in the $\hbar\to 0$ limit, the effective potential is a monotonic function of $\phi$ between $\tilde{\varphi}_-$ and $\tilde{\varphi}_+$:
\ba
\Gamma(\phi)& \stackrel[\hbar\to 0]{=}{} &\frac{(\tilde{\varphi}_+-\phi)\tilde{\Gamma}_-+(\phi-\tilde{\varphi}_-)\tilde{\Gamma}_+}{\tilde{\varphi}_+-\tilde{\varphi}_-}.
\ea
This is the Maxwell construction. To the left of the branch point at $\phi=\tilde{\varphi}_-$ and to the right of the branch point at  $\phi=\tilde{\varphi}_+$, we have only one saddle, at $\tilde{\varphi}$ say, and $\mathcal{J}(\phi)=V'(\phi)$ (to zeroth order in $\hbar$). For the case with $V(\Phi)=-\Phi^2/2+\Phi^4/4$, we have $\tilde{\varphi}_+=-\tilde{\varphi}_-\equiv\tilde{\varphi}$ and $\Gamma(\phi)=\tilde{\Gamma}$ for $-\tilde{\varphi}<\phi<\tilde{\varphi}$. The similarity of the above zero-dimensional result for the Maxwell construction with the higher-dimensional field-theory case is presented for completeness in appendix~\ref{app:B}.

%%%%%%%%%%%%%%%%%%%%%%%%%%%%%%%%%%%%%%%%%%%%%%%%%%%%%
\section{Conclusion}
\label{sec:Conclusions}
%%%%%%%%%%%%%%%%%%%%%%%%%%%%%%%%%%%%%%%%%%%%%%%%%%%%%%

We have provided an explicit exposition of the two-particle irreducible (2PI) effective action for a zero-dimensional quantum field theory. In doing so, we have been able to clarify in detail the behaviour of the sources, and the relationships between the variables of the Legendre transform and the saddle points of the path integral. Moreover, we have confirmed the self-consistency of the approach first presented in \cite{Garbrecht:2015cla}, wherein it was shown that the sources can be used consistently to drive the saddle point of the path integral towards the physical quantum-corrected configuration, providing an improved perturbation theory. Finally, we have explicitly illustrated the convexity of the 2PI effective action and clarified subtle details of the Maxwell construction (with respect to the implicit dependencies on the convex-conjugate variables) in the case of two competing saddle points. The analysis presented here generalises straightforwardly to higher PI effective actions (see, e.g., \cite{Carrington:2004sn}), where one has the additional freedom of higher-order sources (coupling to higher powers of the field). In a future work, we will present similar zero-dimensional considerations in the case of models with global symmetries and involving anticommuting variables.

\appendix

\section{Unstable saddle}
\label{app:A}

In order to see that the contribution from the central, unstable saddle point is negligible, we consider the corresponding integral
\ba
Z_0({\cal J},{\cal K})&=&\exp\left[ -\frac{1}{\hbar}\left( S(\varphi_0)-{\cal J}\varphi_0-\frac{1}{2}{\cal K}\varphi_0^2 \right) \right]\nonumber\\
	&~&\times \int{\rm d}\hat\Phi_0\exp\left[ -\frac{\hbar^{1/2}}{3!}\lambda\varphi_0\hat\Phi_0^3-\frac{\hbar}{4!}\lambda\hat\Phi_0^4 \right]\exp\left[+\frac{1}{2}|{\cal G}_0^{-1}|\hat\Phi_0^2 \right].\nonumber\\
\ea
While the quadratic term is now positive, the integral nevertheless converges thanks to the $\hat{\Phi}^4_0$ term. Since the integral is convergent, the additional exponential suppression of the contribution from $\varphi_0$ relative to $\varphi_{\pm}$ (due to its larger source-dependent action) is sufficient to see why the central saddle point can be neglected (cf.~figure~\ref{fig:saddle_contribution}). The remaining integral has three saddle points itself, and these are given by
\ba
\xi_0=0\quad {\rm and}\quad \xi_{\pm}=-\frac{3}{2}\frac{\varphi_0}{\hbar^{1/2}}\pm\frac{\sqrt{3}}{2\hbar^{1/2}\lambda^{1/2}}\left(8|\mathcal{G}_0^{-1}|+3\lambda\varphi_0^2\right)^{1/2},
\ea
satisfying
\ba
|\mathcal{G}_0^{-1}|\xi_i-\frac{\hbar^{1/2}}{2}\lambda \varphi_0\xi_i^2-\frac{\hbar}{6}\lambda\xi_i^3=0.
\ea
Notice that the two stable saddle points $\xi_{\pm}$ are non-perturbative in $\hbar$. 

\section{Isolating the zero mode}
\label{app:B}

In the case of a multi-dimensional field theory, there is an additional subtlety when we sum over competing saddle points in order to obtain the Maxwell construction. Consider the expression for the one-point function in (\ref{eq:onepointMax}):
\ba
\label{eq:phiapprox}
\phi\approx\frac{\varphi_-Z_-+\varphi_+Z_+}{Z_-+Z_+}.
\ea
Since the eigenspectra of fluctuations around the saddle points $\varphi_{\pm}$ are, in general, distinct, disconnected vacuum diagrams cannot cancel in the ratios $Z_{\pm}/(Z_-+Z_+)$, appearing in (\ref{eq:phiapprox}), as they do in the exact expression for $\phi$. However, in the case of the Maxwell construction, we are interested only in the zero mode, corresponding to a homogeneous configuration, and the resolution to this problem is to partition unity so as to project out only this contribution. A lucid discussion of this in the case of finite-temperature phase transitions is presented in \cite{Rivers:1983sq}, and, for completeness, we review the key details below, generalising for the inclusion of the bilocal source $\mathcal{K}[x,y;\phi,\Delta]$.

Working in four-dimensional Euclidean space, we isolate the zero-momentum component of $\Phi(x)$ by partitioning unity in the form
\ba
1\ =\ \int_{-\infty}^{+\infty}{\rm d} \mathcal{\varphi}\;\delta\left(\varphi-\Omega^{-1}\int\!{\rm d}^4x\;\Phi(x)\right),
\ea
where $\Omega$ is the Euclidean four-volume. Inserting this into the partition function, we have
\ba
Z[\mathcal{J},\mathcal{K}]&\propto&\int\mathcal{D}\Phi\int_{-\infty}^{+\infty}{\rm d}\varphi\;\delta\left(\varphi-\Omega^{-1}\int\!{\rm d}^4x\;\Phi(x)\right)\nonumber\\&&\times\exp\left[-\frac{1}{\hbar}\left(S[\Phi]-\int\!{\rm d}^4x\;\mathcal{J}[x;\phi,\Delta]\Phi(x)\right.\right.\nonumber\\&&-\left.\left.\frac{1}{2}\int\!{\rm d}^4x\int\!{\rm d}^4y\;\Phi(x)\mathcal{K}[x,y;\phi,\Delta]\Phi(y)\right)\right],
\ea
where $\mathcal{D}\Phi$ is now a \emph{functional} measure, and $\mathcal{J}[x,y;\phi,\Delta]$ and $\mathcal{K}[x,y;\phi,\Delta]$ are now \emph{functionals} of the one- and two-point functions $\phi$ and $\Delta$. Throughout this appendix, we use $\propto$ to indicate that we are ignoring the overall constant normalisation of the path integral (constant with respect to the parameters of the theory). We now expand the integrand by decomposing $\Phi(x)=\varphi+\sqrt{\hbar}\hat{\Phi}(x)$. We then obtain
\ba
\label{eq:linearstep}
Z[\mathcal{J},\mathcal{K}]&\propto&\int_{-\infty}^{+\infty}{\rm d}\varphi\;\exp\left[-\frac{\Omega}{\hbar}\left(S(\varphi)-\mathcal{J}[\phi,\Delta]\varphi-\frac{1}{2}\mathcal{K}[\phi,\Delta]\varphi^2\right)\right]\nonumber\\&&\times \int\mathcal{D}\hat{\Phi}\;\delta\left(\int\!{\rm d}^4x\;\hat{\Phi}(x)\right)\nonumber\\&&\times\exp\left[-\frac{1}{2}\int\!{\rm d}^4x\int\!{\rm d}^4y\;\hat{\Phi}(x)\mathcal{G}^{-1}(x,y;\varphi)\hat{\Phi}(y)\right]\nonumber\\&&\times\exp\left[-\frac{1}{\hbar^{1/2}}\int\!{\rm d}^4x\;\hat{\Phi}(x)\left(\frac{\delta S[\Phi]}{\delta\Phi}\bigg|_{\Phi=\varphi}-\mathcal{J}[x;\phi,\Delta]\right.\right.\nonumber\\&&\left.\left.-\int\!{\rm d}^4y\;\mathcal{K}[x,y;\phi,\Delta]\varphi\right)\right]\left[1+\mathcal{O}(\hbar^{1/2})\right],
\ea
where
\numparts
\ba
\mathcal{J}[\phi,\Delta]&\equiv& \Omega^{-1}\int\!{\rm d}^4x\;\mathcal{J}[x;\phi,\Delta]\;,\\
\mathcal{K}[\phi,\Delta]&\equiv& \Omega^{-1}\int\!{\rm d}^4x\int\!{\rm d}^4y\;\mathcal{K}[x,y;\phi,\Delta],
\ea
\endnumparts
and
\numparts
\ba
\label{eq:Gfuncdef}
\mathcal{G}^{-1}(x,y;\varphi)&=&G^{-1}(x,y;\varphi)-\mathcal{K}[x,y;\phi,\Delta]\;,\\ G^{-1}(x,y;\varphi)&=&\frac{\delta^2 S[\Phi]}{\delta \Phi(x)\delta \Phi(y)}\bigg|_{\Phi=\varphi}.
\ea
\endnumparts
In addition, we have defined the notation $S(\varphi)$ via $\Omega S(\varphi)\equiv S[\varphi]$ for constant $\varphi$. Note that
\ba
\label{eq:divdiffs}
S^{(2)}(\varphi)=\frac{\partial^2S(\varphi)}{\partial \varphi^2}\neq \frac{\delta^2 S[\Phi]}{\delta \Phi(x)\delta \Phi(y)}\bigg|_{\Phi=\varphi}.
\ea

If we restrict to translationally invariant situations then $\mathcal{J}[x;\phi,\Delta]$ is constant with respect to $x$ and $\mathcal{K}[x,y;\phi,\Delta]$ depends, at most, on the relative coordinate $x-y$. In this case, we can write
\ba
\int\!{\rm d}^4x\int\!{\rm d}^4y\;\hat{\Phi}(x)\mathcal{K}[x,y;\phi,\Delta]&=&\int\!{\rm d}^4x\int\!{\rm d}^4(x-y)\;\hat{\Phi}(x)\mathcal{K}[x-y,0;\phi,\Delta]\nonumber\\ &=& \mathcal{K}[\phi,\Delta]\int\!{\rm d}^4x\;\hat{\Phi}(x),
\ea
and the linear terms in $\hat{\Phi}$ in the fourth and fifth lines of (\ref{eq:linearstep}) are removed by the constraint
\ba
\int\!{\rm d}^4x\;\hat{\Phi}(x)=0,
\ea
i.e.~that the spacetime average of the fluctuations is zero. We are then left with
\ba
Z[\mathcal{J},\mathcal{K}]&\propto&\int_{-\infty}^{+\infty}{\rm d}\varphi\;\exp\left[-\frac{\Omega}{\hbar}\left(S(\varphi)-\mathcal{J}[\phi,\Delta]\varphi-\frac{1}{2}\mathcal{K}[\phi,\Delta]\varphi^2\right)\right]\nonumber\\&&\times \int\mathcal{D}\hat{\Phi}\;\delta\left(\int\!{\rm d}^4x\;\hat{\Phi}(x)\right)\nonumber\\&&\times\exp\left[-\frac{1}{2}\int\!{\rm d}^4x\int\!{\rm d}^4y\;\hat{\Phi}(x)\mathcal{G}^{-1}(x,y;\varphi)\hat{\Phi}(y)\right]\left[1+\mathcal{O}(\hbar)\right],\nonumber\\
\ea

We now proceed by rewriting the delta function as an integral over an auxiliary parameter $\xi$ via
\ba
\delta\left(\int\!{\rm d}^4x\;\hat{\Phi}(x)\right) \propto \int_{-\infty}^{+\infty}{\rm d}\xi\;\exp\left[i\xi\int\!{\rm d}^4x\;\hat{\Phi}(x)\right],
\ea
such that
\ba
Z[\mathcal{J},\mathcal{K}]&\propto&\int_{-\infty}^{+\infty}{\rm d}\varphi\;F(\varphi)\exp\left[-\frac{\Omega}{\hbar}\left(S(\varphi)-\mathcal{J}[\phi,\Delta]\varphi-\frac{1}{2}\mathcal{K}[\phi,\Delta]\varphi^2\right)\right],\nonumber\\
\ea
with
\ba
F(\varphi)&\propto& \int_{-\infty}^{+\infty}{\rm d}\xi\int\mathcal{D}\hat{\Phi}\;\exp\left[-\frac{1}{2}\int\!{\rm d}^4x\int\!{\rm d}^4y\;\hat{\Phi}(x)\mathcal{G}^{-1}(x,y;\varphi)\hat{\Phi}(y)\right.\nonumber\\&&\left.+i\xi\int\!{\rm d}^4x\;\hat{\Phi}(x)\right]\left[1+\mathcal{O}(\hbar)\right].
\ea
Performing the functional integral, we have
\ba
F(\varphi)&\propto& \int_{-\infty}^{+\infty}{\rm d}\xi\;{\rm det}^{-1/2}\left[\mathcal{G}^{-1}(\varphi)\ast G(0)\right]\nonumber\\&&\times\exp\left[-\frac{\Omega}{2}\xi^2\left(S^{(2)}(\varphi)-\mathcal{K}[\phi,\Delta]\right)^{-1}\right]\left[1+\mathcal{O}(\hbar)\right],
\ea
where $\ast$ denotes a convolution, and the remaining $\xi$ integral yields
\ba
F(\varphi)&\propto& \left(S^{(2)}(\varphi)-\mathcal{K}[\phi,\Delta]\right)^{1/2}{\rm det}^{-1/2}\left[\mathcal{G}^{-1}(\varphi)\ast G(0)\right]\left[1+\mathcal{O}(\hbar)\right].
\ea
Thus, we arrive at the expression
\ba
Z[\mathcal{J},\mathcal{K}]&\propto&\int_{-\infty}^{+\infty}{\rm d}\varphi\;\left(S^{(2)}(\varphi)-\mathcal{K}[\phi,\Delta]\right)^{1/2}\nonumber\\&&\times\exp\left[-\frac{\Omega}{\hbar}\left(S(\varphi)-\mathcal{J}[\phi,\Delta]\varphi-\frac{1}{2}\mathcal{K}[\phi,\Delta]\varphi^2\right.\right.\nonumber\\&&\left.\left.+\frac{\hbar}{2\Omega}\ln{\rm det}\left[\mathcal{G}^{-1}(\varphi)\ast G(0)\right]\right)\right]\left[1+\mathcal{O}(\hbar)\right].
\ea
We emphasise that $\mathcal{G}^{-1}(\varphi)\neq S^{(2)}(\varphi)-\mathcal{K}[\phi,\Delta]$, unlike in the zero-dimensional case, by virtue of (\ref{eq:Gfuncdef}) and (\ref{eq:divdiffs}).

Supposing that we now have two relevant saddles $\varphi_{\pm}$ (for which $S^{(2)}(\varphi_{\pm})-\mathcal{K}[\phi,\Delta]>0$), we expand $\varphi=\varphi_{\pm}+\sqrt{\hbar}\hat{\varphi}_{\pm}/\Omega^{1/2}$, giving
\ba
Z[\mathcal{J},\mathcal{K}]&\sim&\sum_{\pm}\left(S^{(2)}(\varphi_{\pm})-\mathcal{K}[\phi,\Delta]\right)^{1/2}\nonumber\\&&\times\exp\left[-\frac{\Omega}{\hbar}\left(S(\varphi_{\pm})-\mathcal{J}[\phi,\Delta]\varphi_{\pm}-\frac{1}{2}\mathcal{K}[\phi,\Delta]\varphi_{\pm}^2\right.\right.\nonumber\\&&\left.\left.+\frac{\hbar}{2\Omega}\ln{\rm det}\left[\mathcal{G}^{-1}(\varphi_{\pm})\ast G(0)\right]\right)\right]\nonumber\\&&\times\int_{-\infty}^{+\infty}{\rm d}\hat{\varphi}_{\pm}\;\exp\left[-\frac{1}{2}\left(S^{(2)}(\varphi_{\pm})-\mathcal{K}[\phi,\Delta]\right)\hat{\varphi}_{\pm}^2\right]\left[1+\mathcal{O}(\hbar)\right].\nonumber\\
\ea
We see that the Gaussian fluctuations integrate to unity and, in isolating the zero mode and dealing with the functional integrals, we have been left with the zero-dimensional field theory of the zero mode, consistent with what we obtained in section~\ref{sec:MultiSaddle}:
\ba
Z[\mathcal{J},\mathcal{K}]&\sim&\sum_{\pm}\exp\left[-\frac{\Omega}{\hbar}\left(S(\varphi_{\pm})-\mathcal{J}[\phi,\Delta]\varphi_{\pm}-\frac{1}{2}\mathcal{K}[\phi,\Delta]\varphi_{\pm}^2\right.\right.\nonumber\\&&+\left.\left.\frac{\hbar}{2}\ln{\rm det}\left[\mathcal{G}^{-1}(\varphi_{\pm})\ast G(0)\right]\right)\right]\left[1+\mathcal{O}(\hbar)\right],
\ea
the exception being the dependence on the volume $\Omega$, such that the Maxwell construction arises in the sequence of limits $\Omega\to\infty$, $\hbar\to 0^+$.

%%%%	

\ack

PM would like to thank Bj\"{o}rn Garbrecht for earlier collaboration on this topic, as well as Wen-Yuan Ai and Jean Alexandre for interesting discussions. This work was supported in part by a Leverhulme Trust Research Leadership Award, and the Science and Technology Facilities Council (STFC) under Grant Nos.~ST/L000393/1 and ST/P000703/1.

%%%%

\end{document}